\documentclass[graybox,natbib,nosecnum]{svmult}
\bibpunct{(}{)}{;}{a}{}{,} 

\pdfoutput=1   

\usepackage{mathptmx}       
\usepackage{helvet}         
\usepackage{courier}        
\usepackage{type1cm}        

\usepackage{makeidx}         
\usepackage{graphicx}        
\usepackage{multicol}        
\usepackage[bottom]{footmisc}
\usepackage[normalem]{ulem}	
\usepackage{hyperref}  
\usepackage{amsmath}
\usepackage{amssymb}
\usepackage{soul}   
\graphicspath{{./}}

\makeindex             

\begin{document}

\title*{Finding planets via gravitational microlensing}
\author{Natalia E. Rektsini \inst{1,2} \& Virginie Batista \inst{2}}

\institute{$^1$ School of Natural Sciences, University of Tasmania,  Private Bag 37 Hobart, Tasmania, 7001, Australia \\ $^2$ Sorbonne Universit\'e, CNRS, Institut d'Astrophysique de Paris, IAP, F-75014, Paris, France 
\\ \email{efstathia.rektsini@utas.edu.au}}

%
%
\maketitle

\abstract{
Since the first microlensing planet discovery in 2003, more than 200 planets have been detected with gravitational microlensing, in addition to several free-floating planet and black hole  candidates. In this chapter the microlensing theory is presented by introducing the numerical methods used to solve binary and triple lens problems and how these lead to the characterisation of the planetary systems. Then the microlensing planetary detection efficiency is discussed, with an emphasis on cold planets beyond the snow line. Furthermore, it will be explained how the planetary characterisation can be facilitated when the microlensing light curves exhibit distortions due to second order effects such as parallax, planetary orbital motion, and extended source. These second order effects can be turned to our advantage, and become useful to ultimately better characterise the planetary systems, but they can also introduce degeneracies in the light curve models. It will be explained how the use of modern observational and computational techniques enables microlensers to solve these degeneracies and estimate the planetary system parameters with very high accuracy. Then a review of the main discoveries to date will be presented while exploring the recent statistical results from high-cadence ground-based surveys and space-based observations, especially on the planet mass function. Finally, future prospects are discussed, with the expected advances from dedicated space missions, extending the planet sensitivity range down to Mars mass.  }

\section{Introduction }

Gravitational microlensing is a unique technique to discover exoplanets because it doesn't rely on the host star brightness or the orbital motion of the planetary system. In contrast to other methods, it relies on time-variable magnification of the light of an unrelated background star due to deflection by the gravitational potential of the host star$+$planet system. Detectable magnification requires very precise alignment of the planetary system with a background source star and the observer. When this alignment occurs, the mass of the intermediate planetary system will act as a lens, deflecting the light emitted by the background source star. As a result, the observer receives light from multiple projected images of the source star, each of which has its own time-variable angular size, which can be interpreted as a time-variable light magnification. 

Predicting exactly where and when a microlensing event will occur as well as how significant the magnification will be is impossible, and the duration of the event is limited and unique. Therefore, the expectation of observing this phenomenon received a rather pessimistic reaction from the scientific community, including Einstein, who was the first to bring attention to it \citep{einstein36}, after receiving multiple requests from R. W. Mandl.

Indeed, millions of stars need to be monitored in order to catch a handful of magnified ones. In the following decades, several scientists considered experiments of microlensing observations \citep[e.g][]{refsdal64, chang79}, and finally \cite{paczynski86} managed to convince the international community to develop the MACHO, EROS and OGLE experiments to search for compact objects in the galactic halo (see acronyms at the end of the chapter). Hunting for microlensing events towards the galactic center with the intention to discover planets appeared a few years later, under the suggestion of \cite{mao91} and \cite{gould92}, who gave a detailed description of microlensing planetary perturbations. The development of this technique over the last 30 years required tremendous efforts and steady determination from the scientific community. 

Nowadays, thousands of microlensing events are observed per year, leading to the detection of tens of exoplanets for each season, accounting for more than 200 planetary systems published to date and many more under analysis. This number is quite modest compared to the total amount of discovered exoplanets but this will change with the launch of the {\it{Nancy Grace Roman Space Telescope}} \citep{spergel15} scheduled in late 2026. {\it{Roman}} will observe 2 square degrees close to the Galactic center with high optical depth to microlensing during 5 observing runs of 72 days each. It is estimated that it will detect more than 1400 bound microlensing exoplanets down to the mass of Mars \citep{Penny2019}, with orbits ranging from 0.5 AU to wide orbits. Given the recent results of \cite{Sumi2023}, we can also expect thousand or more unbound planets down to Mars-masses.

Microlensing offers a considerable advantage in probing a parameter space that is not accessible to other methods. It can detect planets down to Earth masses\citep{bennett96}, within a range of separations to the host star that is complementary to other techniques with maximum sensitivity to planets orbiting close to or beyond the snow line. 
According to the core accretion theory \citep{ida04}, this is where planetary formation is predicted to be the most efficient. Moreover, as the only effect measured by microlensing is that of deflecting light by gravitational distortions of a given field, it is independent of the emitted light of the lens. This makes it unbiased by the lens properties and opens a domain of exploration down to very faint, even invisible objects, such as brown dwarfs, 
black holes, neutron stars and free-floating planets while also making possible the detection of exoplanets in very large distances throughout our Galaxy.

Detecting distant exoplanets means that they are hard or impossible to re-observe so studying their atmospheric and interior properties is not likely. The importance of these detections lies in their implications for theoretical and statistical exoplanet studies. Generally, the formation processes are influenced by many environmental factors, such as stellar metallicity \citep{wang15, zhu16}, as well as the mass of the host star \citep{johnson10}, the star multiplicity \citep{kaib13}, and larger scale conditions like higher ambient radiation and stellar density in the galactic bulge \citep{thompson13}. Thus, the distant population of planets unveiled by microlensing provides critical constraints for planetary formation theories and population synthesis models.  

\section{Observations}
The first microlensing surveys debuted in the early 90's, searching for the existence of dark matter in the galactic halo \citep{Alcock2000,Tisserand2007, wyrzykowski2011}. The idea to use this technique to detect exoplanets was introduced later by \cite{mao91} and  \cite{gould92}. The first generation survey telescopes were able to observe only a small fraction of the sky with a limited observing cadence which restricted their ability to detect the planetary perturbation. A solution to this problem came when the OGLE-early-warning system was introduced announcing alerts of potential microlensing events in real time. This permitted other surveys with multiple smaller telescopes to follow-up on the events. Using numerous observatories at all longitudes and different wavebands ensured a round-the-clock coverage of the planetary anomalies. 

The new millennium marked a real advance in the field, with two microlensing collaborations: MOA \citep{bond01, sumi13} and OGLE \citep{udalski03} in addition to the  follow-up collaborations: PLANET \citep{Albrow1998}, $\mu$FUN \citep{Gould2010}, RobotNet \citep{Tsapras2009} and MiNDSTEp \citep{Dominik2010}. The Wise observatory in Israel also conducted a survey \citep{gorbikov10}, despite its less favorable northern latitude. All this hard effort and team work lead to the first microlens exoplanet discovery \citep{bond04} and to hundred more detections ever since, establishing microlensing as a successful exoplanet detection technique.

Almost a decade later, the microlensing community made a transition towards a network of wide field imagers. The second generation surveys conduct high cadence observations from different longitudes of the southern hemisphere. In 2006, the MOA collaboration started using a 2.2-deg$^2$ FOV CCD camera \citep{sako08} mounted on a 1.8m MOA-II telescope in New-Zealand. The OGLE collaboration upgraded its camera in 2010 to a 1.4-deg$^2$ FOV on a 1.3-m telescope in Chile. Additionally, KMTNet \citep{henderson14,kim16} became operational in 2015, with 3 telescopes equipped with a 4 deg$^2$ camera, in Chile, South Africa and Australia. The addition of this microlensing survey had a significant impact on the efficiency of microlensing planet detections by discovering or contributing to the discovery of 204 out of 278 microlensing planets until today according to their site (accessed 24/3/2024):(\url{https://kmtnet.kasi.re.kr/ulens/data/KMTNet_exoplanet_list.pdf}). These three surveys are continuously monitoring hundreds of fields mostly towards the bulge. Today most of the follow-up groups have stopped operating but some of the observatories (e.g., LCO, UTGO ) are still following alerts from surveys dedicated and not dedicated to microlensing (e.g., ASAS, ZTF, GAIA).

In 2014-2017, the microlensing campaigns from the ground were also combined with space-based observations using the telescopes {\it{Spitzer}} \citep{werner04, gould13a}  and {\it{Kepler}} ({\it{K2-C9}}) \citep{howell14}, as a pathway towards the future space missions {\it{Roman}} \citep{spergel15, Penny2019} and {\it{Euclid}} \citep{penny13}.   

The PRime-focus Infrared Microlensing Experiment (PRIME) started operating in late 2023 \citep{Kondo2023}. It is the first near-infrared microlensing survey, using a 1.8m telescope with a wide field of view of 1.45 deg$^2$, positioned at the South African Astronomical Observatory. The main goal of this survey, amongst others, is to explore and derive optimised observational strategies for {\it{Roman}} ensuring the maximum possible microlensing exoplanet detections towards the Galactic Bulge. 

Since 2016 {\it{GAIA}} \citep{Gaia2016} is using astrometric techniques covering the whole galaxy, the galactic bulge, the plane and the disk at a wide range of magnitudes. Their photometric alert system has proven very valuable for detecting and following-up on microlensing events on any kind of magnitudes, using both photometric and astrometric microlensing \citep{Wyrzykowski2023}.
Ultimately, Gaia will include astrometric time series in their fourth data release (DR4), that will be extremely valuable to  joined photometic and astrometric fits.

Interferometric methods have also been used for microlensing events, with the first resolved microlensing image by VLTI GRAVITY \citep{Dong2019} which was followed by the first images of rotating gravitational microlensing arcs, using the VLTI PIONER instrument \citep{Cassan2022}. These successful observations have opened the path to interferometric microlensing which can potentially lead to high precision mass measurements for future microlensing events. 

In addition, microlensing will also be included in the scientific goals of the Legacy Survey of Space and Time (LSST) survey of the Southern Hemisphere sky \citep{Abrams2023,Street2023b}. The survey will be conducted with the Vera C. Rubin Observatory using an 8.4-meter telescope situated at the top of Cerro Pachon ridge in Chile. In order to prepare for this survey the LCO team have been running the OMEGA (2020-2023) and OMEGA 2 key projects, running all-sky photometric and spectroscopic follow-up for microlensing alerts from MOA, OGLE , GAIA and ZTF. 

Finally, the {\it{Nancy Grace Roman Space Telescope}} will be the first space-based telescope with a survey (RGBTDS) dedicated to the microlensing technique. {\it{Roman}} is a NASA flagship mission scheduled for launch in October 2026 towards the Lagrange point L2. 
The concept was initiated with a dedicated mission \citep{bennett02} called the Microlensing Planet Finder (MPF), which had been proposed to NASA's Discovery program but not selected. The objective was to be able to monitor turn off stars in the galactic bulge as microlensing sources. 
The Roman Wide Field Instrument (WFI) consists of 18 4k$\times$4k Hawaii H4RG-10 HgCdTe detectors with a pixel scale of 0.11" with a 0.28 deg$^2$ field of view \citep{spergel15}. It will also contain 6 broad band filters (R062, Z087, Y106, J129, H158, F184, K213) from 0.62 $\mu$m to 2.13 $\mu$m and an extremely wide filter from 0.92 $\mu$m to 2.13 $\mu$m (W146). {\it{Roman}} will observe 7 fields covering 1.97 square degrees, with a cadence of 15 minutes in W146 for 6 periods of 72 days.

{\it{Euclid}} is an ESA mission, launched in July 2023, with primary goals to measure parameters of dark energy using weak gravitational lensing and baryonic acoustic oscillation, test the general relativity and the Cold Dark Matter paradigm for structure formation. Since its original submission in 2007 to ESA, a microlensing planet hunting program has been listed as part of the Legacy science \citep{beaulieu08}. 
The vision adopted by the Europeans of a joint mission with Dark Energy probes and microlensing has been promoted and adopted by the Astro 2010 Decadal Survey when it created and ranked as top priority the {\it{Roman}} mission.

A baseline for the design of the {\it{Euclid}} microlensing survey was firstly described by \cite{penny13} with a detailed simulation demonstrating 
the capabilities and the expected scientific outcomes. \citep[]{Bachelet2019,Bachelet2022} conducted the same simulations for the {\it{Roman}} survey and predicted that precursor {\it{Euclid}} observations of all the currently envisioned  Roman fields would be very valuable in defining exactly which fields will be observed, would help to understand the structure and extinction of the line of sight, and provide a reference for relative source-lens proper motion of forecoming {\it{Roman}} microlensing events.
The objective is to perform 4 {\it{Euclid}} observations at 9 pointing positions covering 4.5 square degree. The procedure will take 42 hours, which is very a valuable use of {\it{Euclid}} time. Both VIS and IR data will be used. The VIS instrument contains 36 4k$\times$4k e2v CCD273-84 CCD detectors with 0.10" pixels, giving a 0.47 deg$^2$ field of view while the NISP instrument has the same field of view with 16 H2RG-18 HgCdTe detectors with a pixel scale of 0.30".

\section{Microlensing basics}

\begin{figure}[h!]
\begin{center}
\includegraphics[scale=.2,trim=50 240 100 230, angle=-90]{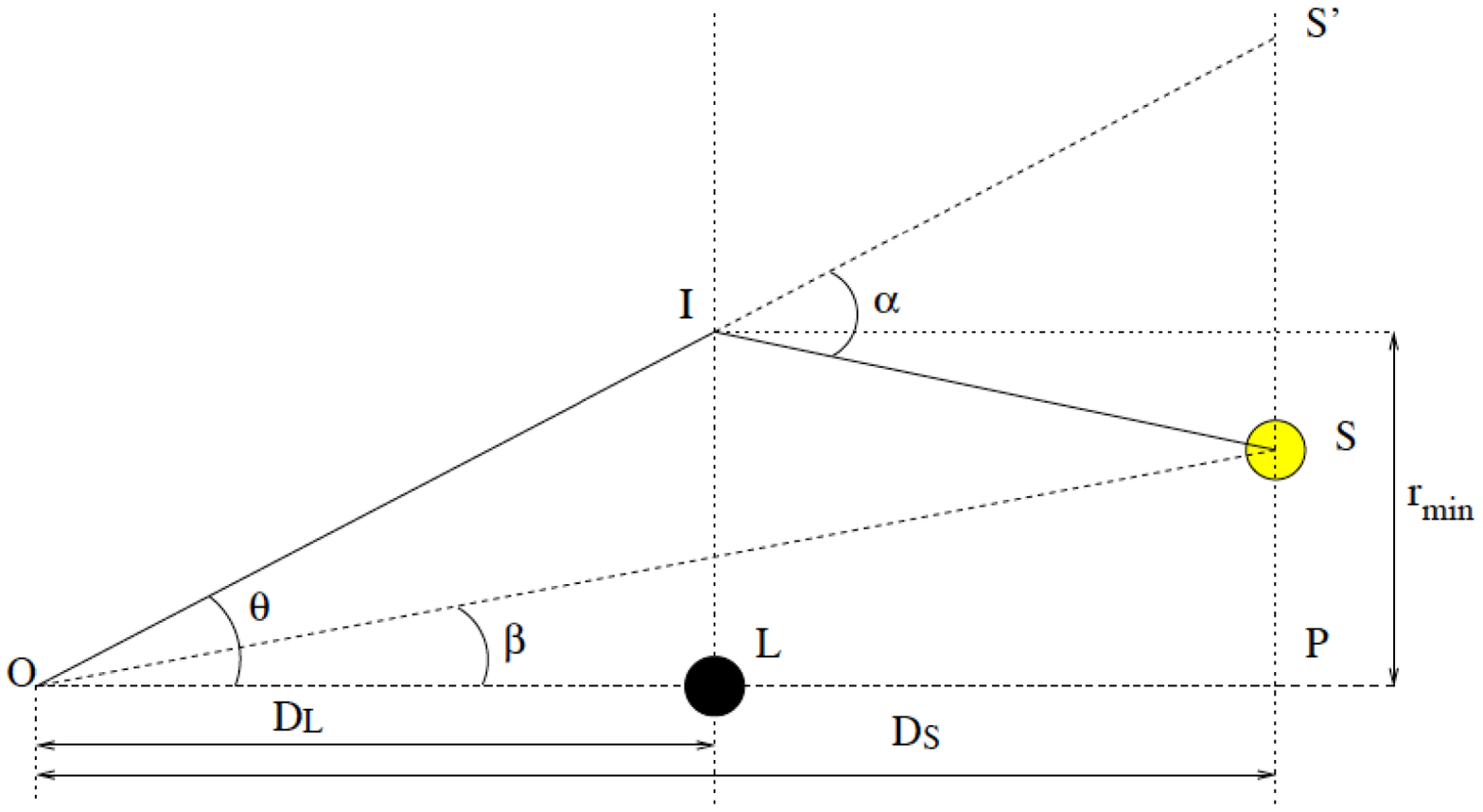}
\caption{Deviation of a light ray from a source S due to the gravitational field of a microlens L.}
\label{fig:1}       
\end{center}
\end{figure}

The basic principle of a microlensing effect is illustrated in Fig.~\ref{fig:1}, showing a light ray coming from a source S being deflected by a point-like mass L, called `lens'. The light is deviated by an angle $\alpha$, creating the illusion for the observer (O) that the source is at the position I (`image') in the lens plane. The three angles of this figure are linked by the lens equation, $\beta=\theta-\alpha$, where $\alpha=4GM/(c^2D_L\theta)$, with $M$ the lens mass, $G$ the gravitational constant and $c$ the speed of light. Under the assumption of small angles, one can write $\alpha(D_S-D_L)=\beta D_S$.

The deviation of light does not affect only one ray but an ensemble of rays coming from the source, so when the source and the lens are perfectly aligned on the observer's line of sight, $\beta=0$ and the source image in the lens plan appears as a ring, called the `Einstein ring', whose angular size $\theta_{\rm{E}}$ is expressed as

\begin{equation}
\theta_{\rm{E}}=\sqrt{\kappa M\frac{D_S-D_L}{D_S D_L}},\quad {\rm{where}}\,\,\kappa=4G/c^2\simeq 8.14 \,\rm{kpc.M}^{-1}_\odot.
\end{equation}

The lens equation is thus commonly expressed as $\beta=\theta-\theta_{\rm{E}}^2/\theta$. The Einstein angular radius $\theta_{\rm{E}}$ is used as a scale standard in microlensing units and as long as its value remains unknown, most microlensing parameters will be expressed in units of $\theta_{\rm{E}}$. 

\subsection{Single lens}

By normalizing the angles by $\theta_{\rm{E}}$ and define the parameters $u=\beta/\theta_{\rm{E}}$ and $z=\theta/\theta_{\rm{E}}$, we can express the lens-source separation $u$ as a function of the separation $z$ between the image of the source and the lens, both being projected in the lens plane, $u=z-1/z$. If the lens and the source are not perfectly aligned ($u\neq0$), this equation has two solutions: $z_\pm=\pm (\sqrt{u^2+4}\pm u)/2$. The positive solution creates the major image, and the negative one the minor image, outside and inside of the Einstein ring, respectively.

The magnification of the source by the lens is then estimated as the area ratio of the images over the source. Since the images are tangentially stretched by a factor $z_\pm /u$ compared to the source and radially shrunk by a factor $dz_\pm /du$, the magnification is then expressed as

\begin{equation}
A_\pm=\left|{\frac{z_\pm}{u}\frac{dz_\pm}{du}}\right|=\frac{1}{2}\left[\frac{u^2+2}{u\sqrt{u^2+4}}\pm 1\right],\quad {\rm{thus}\,\,\rm{in}\,\,\rm{total,}}\quad A(u)=\frac{u^2+2}{u\sqrt{u^2+4}}.
\end{equation}

In its simplest form, the lens-source relative movement is rectilinear and $u$ can be expressed as a function of the impact parameter $u_0$ (in units of $\theta_{\rm E}$), the time of maximum magnification $t_0$, and the Einstein crossing time $t_{\rm E}$

\begin{equation}
u(t)=\left[u^2_0+\left(\frac{t-t_0}{t_{\rm E}}\right)^2\right]^{1/2}.
\end{equation}

The observed flux is a function of time, $F(t) = A(t)F_s+F_b$, where $F_s$ is the source flux and $F_b$ the blended light (any unresolved light) that is not magnified. It reaches its maximum at $t_0$, when the impact parameter is at its minimum $u_0$. The smaller $u_0$ is, the higher the maximum of magnification will be. For a Point-Source-Point-Lens (PSPL) event, the light curve can be fit by five parameters: $t_0$, $u_0$, $t_{\rm{E}}$, $F_s$ and $F_b$. In practice, each telescope has its specific $F_s$ and $F_b$, since different telescopes may have different filter bandpasses and resolutions. 


\subsection{Multiple lens}

Let's consider now a lens composed of $N_L$ objects, which the total mass will be $M=\Sigma^{N_L}_{i=1}m_i$. Let $\mathbf{\theta}_i$ be the 2-dimension coordinates of these individual masses in the lens plane. The deflection angle of the source is then expressed by

\begin{equation}
\mathbf{\alpha}(\mathbf{\theta})=\frac{4G}{c^2}\left(\frac{1}{D_L}-\frac{1}{D_S}\right)\sum_i^{N_L}m_i\frac{\mathbf{\theta}-\mathbf{\theta}_i}{|\mathbf{\theta}-\mathbf{\theta}_i|^2},
\end{equation}

where $\mathbf{\theta}$ are the angular positions of the images. It is common to consider the lens equation in complex coordinates \citep{witt90, rhie97}. The source position $\mathbf{u}=(\xi,\eta)$ and the images position $\mathbf{z}=(x,y)$ can then be defined in the complex form, $u=\xi+i\eta$ and $z=x+iy$ respectively. The lens equation thus becomes
\begin{equation}
u=z-\sum_i^{N_L}\frac{m_i/M}{\bar{z}-\bar{z}_i},
\end{equation}
where $\bar{z}$ is the complex conjugate of $z$. Eq.(5) can be solved numerically to find the images position $z_j$, where $j$ is the index for the source images and $i$ those of the individual lens masses. The magnification can be greater or less than 1, depending on the images area, and is equal to the inverse of the Jacobian matrix determinant,

\begin{equation}
A_j=\frac{1}{\rm{detJ}}\Big|_{z=z_j},\quad {\rm{detJ}}\equiv \frac{\partial(\xi,\eta)}{\partial(x,y)}=1-\frac{\partial u}{\partial z}\frac{\overline{\partial u}}{\partial z}.
\end{equation}

The total magnification is the sum of individual ones: $A=\Sigma_j|A_j|$. Singularities occur for positions of the source where ${\rm{detJ}}=0$, theoretically inducing an infinite magnification in the case of a point source. In practise, the source cannot be considered as a point in most cases and the singularities appear as very high but finite magnifications. 

From Eq.(5), $\,\partial u / \partial \bar{z}=\sum\left[\epsilon_i/(\bar{z_i}-\bar{z})^2\right]$, where $\epsilon_i=m_i/M$ is the mass fraction of individual masses $m_i$. According to Eq.(6), the image positions for which ${\rm{detJ}}=0$ are given by

\begin{equation}
\left|\sum_i^{N_L}\frac{\epsilon_i}{(\bar{z_i}-\bar{z})^2}\right|^2=1.
\end{equation}

The solutions of Eq.(7) shape as closed contours in the lens plane, called `critical curves'. To these curves correspond an ensemble of source positions in the source plane, called `caustics'. 


When the lens is composed of two masses, $N_L=2$, the lens equation is expressed by

\begin{equation}
u=z+\frac{\epsilon_1}{\bar{z_1}-\bar{z}}+\frac{\epsilon_2}{\bar{z_2}-\bar{z}}.
\end{equation}

The Eq.(8) can be written as a 5-degree polynomial, whose coefficients are given in \cite{witt95}. The solutions to this polynomial are not necessarily solutions to the Eq.(8), because when the source is outside of the caustics, two of the five solutions yield to an imaginary magnification. Thus, the number of images in the lens plane varies from three to five when the source crosses a caustic.

\begin{figure}
\begin{center}
\includegraphics[angle=-90,scale=.2,trim=0 0 0 0, clip=true]{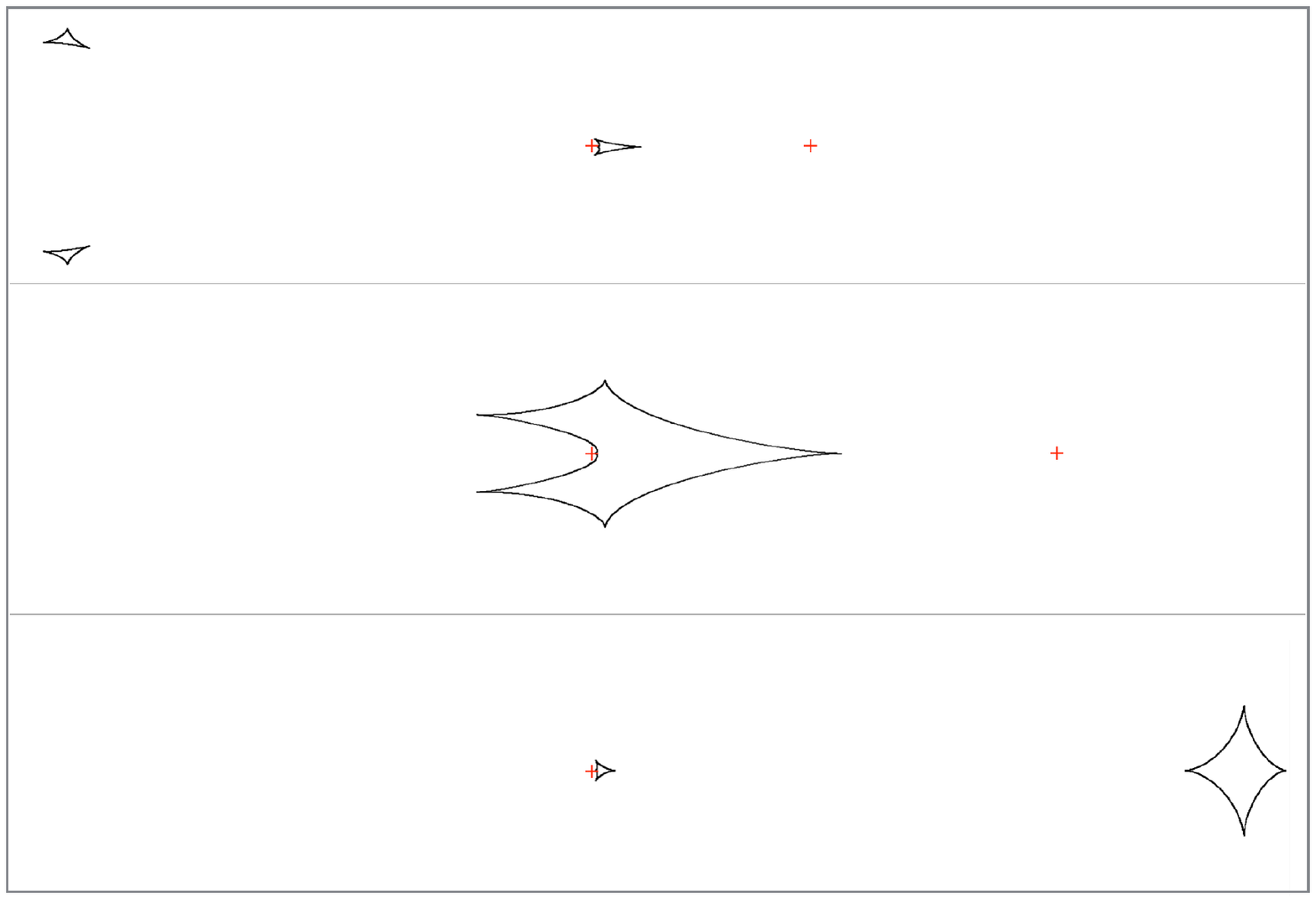}
\caption{Topology of caustics for three regimes of separation, `close' (top), `resonant' (middle) and `wide' (bottom).  }
\label{fig:2}       
\end{center}
\end{figure}

\cite{schneider86} showed that there are three different topologies of caustics for a given value of mass ratio $q$ between the two components of the lens. These three topologies present either one, two or three caustics. As shown on Fig.~\ref{fig:2}, in the case of small separations between the two lenses, $d\ll 1$ (in Einstein angular radius units), there are three caustics, one being the central caustic (with 4 cusps) and two secondary ones (with 3 cusps). The transition towards big separations goes through an intermediary phase that creates a single caustic with 6 cusps, of bigger size, called `resonant caustic'. This configuration happens when the separation is close to the Einstein radius. Finally, for large separations, this caustic splits into two caustics, with 4 cusps each. Naturally these three domains are called `close', 'resonant' and `wide' separations.

For triple lens modeling, \cite{rhie02} gives the lens equation for point source calculations. It is solved the same way as the binary lens equation, but the polynomial is of 10th order instead of 5th. 

\subsubsection{Lensing degeneracies}

\begin{figure}
\begin{center}
\includegraphics[angle=-90,scale=.3,trim=100 0 100 0, clip=true]{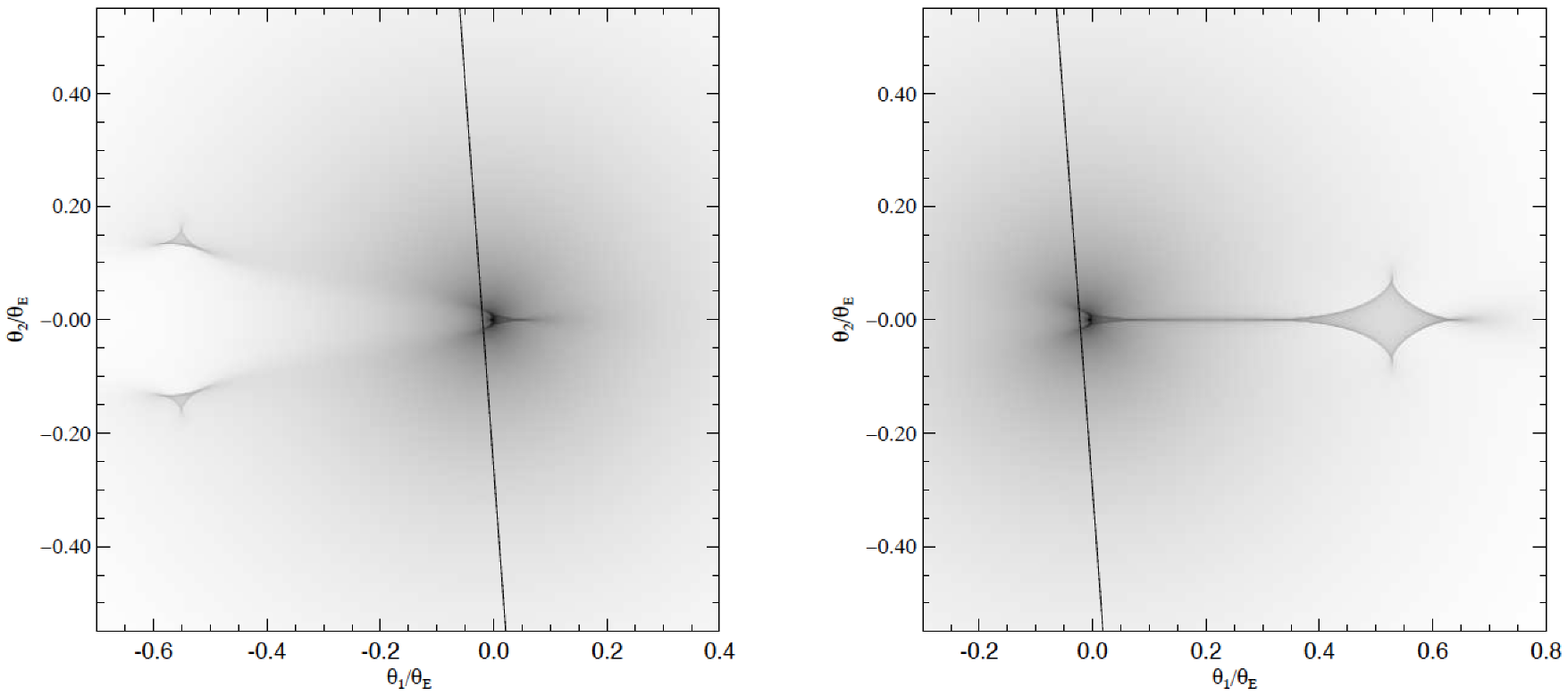}
\caption{Example of $d\longleftrightarrow d^{-1}$ degeneracy. }
\label{fig:3}       
\end{center}
\end{figure}

For mass ratios $q \ll 1$ and a given separation $d$, the shape and size of the central caustic is invariant under the $d\longleftrightarrow d^{-1}$ transformation. This duality is often called `close-wide degeneracy' \citep{griest98, dominik99, albrow99, an05}. It comes from the fact that the Taylor development of the lens equation to the second degree is identical for $d\longleftrightarrow d^{-1}(1+q)^{1/2}$, i.e. $d \longleftrightarrow d^{-1}$ when $q\ll 1$. Fig.~\ref{fig:3} gives an illustration of this degeneracy, where the source passes very close to the central caustic of a system with a $q=0.006$ mass ratio. Two ($d,1/d$) models provide the same light curve. This figure also points out the importance of high magnification events, for which $u_0\ll1$, because when the source passes extremely close to the central caustic, it probes the effect of the planetary companion on the caustic shape.

The second known degeneracy occurs when the source passes close to the planetary caustic. The Chang–Refsdal
\citep{chang79} approximation which is used to describe the planetary caustics \citep{Gaudi1997, dominik98} describes a point-mass lens with uniform shear. This generates a symmetry along the star-planet axis and the line perpendicular to this axis, causing identical light-curves for inner-position or outer-position of the planet, named as the inner-outer degeneracy. 

First, \cite{Yee2021} discovered inconsistencies in the formalism of the close-wide degeneracy suggesting a connection between this two-fold degenerate solution (close-wide and inner-outer degeneracies). On this thought, \cite{Zhang2022} used machine learning techniques to explore in great length these lensing degeneracies, testing various separations and models in a fast and rigorous way. They used a Neural Density Estimator (NDE) capable to learn how to produce posteriors of the lens parameters from a large and complex sample of microlensing light curves. They trained the network using 691.257 simulated microlensing events for {\it{Roman}}. The use of NDE network permits them to retrieve fast and automatically the posterior distributions of a large and rich sample of binary-lens events. The study of this huge variety of light curves showed that there is no real caustic degeneracy and that the close-wide and inner-outer degenerate solutions can be seen as two cases of a more generalised, unified magnification degeneracy. They call this the offset degeneracy and they introduce an analytical formalism to demonstrate this magnification degeneracy \citep{Zhang2022b}. They also revisited 23 published microlensing events that produced this two-fold degenerate solution and prove that their unified offset degeneracy is valid and can be applied in all the events. Finally, \cite{Zhang2022b} use the superposition principle \citep{Bozza1999, Han2001} and deduce a generalised N-body offset degeneracy of a 2$^{N-1}$ number of degenerate configurations.

A third form of lensing degeneracy, defined  as “central caustic cusp
approach” degeneracy \citep{Terry2022} was detected on the OGLE-2011-BLG-0950 event \citep{Choi2012}, where 
the light curve perturbation can be interpreted both as a planetary or stellar binary anomaly. \cite{Terry2022} conducted a follow-up analysis of this event using AO images of the target that permitted them to resolve this sort of degeneracy by constraining the source-lens relative proper motion and finally, to prove the target to be a stellar binary  system. Breaking this form of degeneracy is quite critical  because, contrary to the previously discussed degeneracies, even a well sampled dataset of the perturbation will probably not help to determine this double interpretation.

\subsection{Finite source effects and computation}

As previously mentioned, in reality the source is a disk and the point-like approximation is only valid when the source is far enough from the lens (projected in the lens plane). This means that when the impact parameter is very small, $u_0\ll 1$, finite source effects have to be taken into account at the peak of the microlensing event, and require specific numerical treatments. It is also true when the source approaches or crosses a caustic in the case of multiple lens systems. In the first case, the corresponding effect on the light curve peak is very noticeable, with a rounded shape and a damped magnification. This effect was first analyzed by \cite{gould94} and \cite{nemiroff94}. They showed that the deviation from the standard PSPL light curve provides information on the source proper motion, under assumptions on its brightness and color. Indeed, one can determine $\theta_{\rm{E}}=\theta_*/\rho$, where $\rho$ is the source radius in Einstein units, and $\theta_*$ the physical source size in $\mu$as. The first parameter is derived from the light curve fit, and the latter from available information on the source star color \citep{bensby13} and brightness \citep{nataf13}, and empirical color/brightness relations \citep{kervella04}. Thus, the lens-source relative proper motion is derived by $\mu=\theta_{\rm{E}}/t_{\rm{E}}$. 

\begin{figure}
\begin{center}
\includegraphics[scale=.2,trim=0 150 0 150, clip=true]{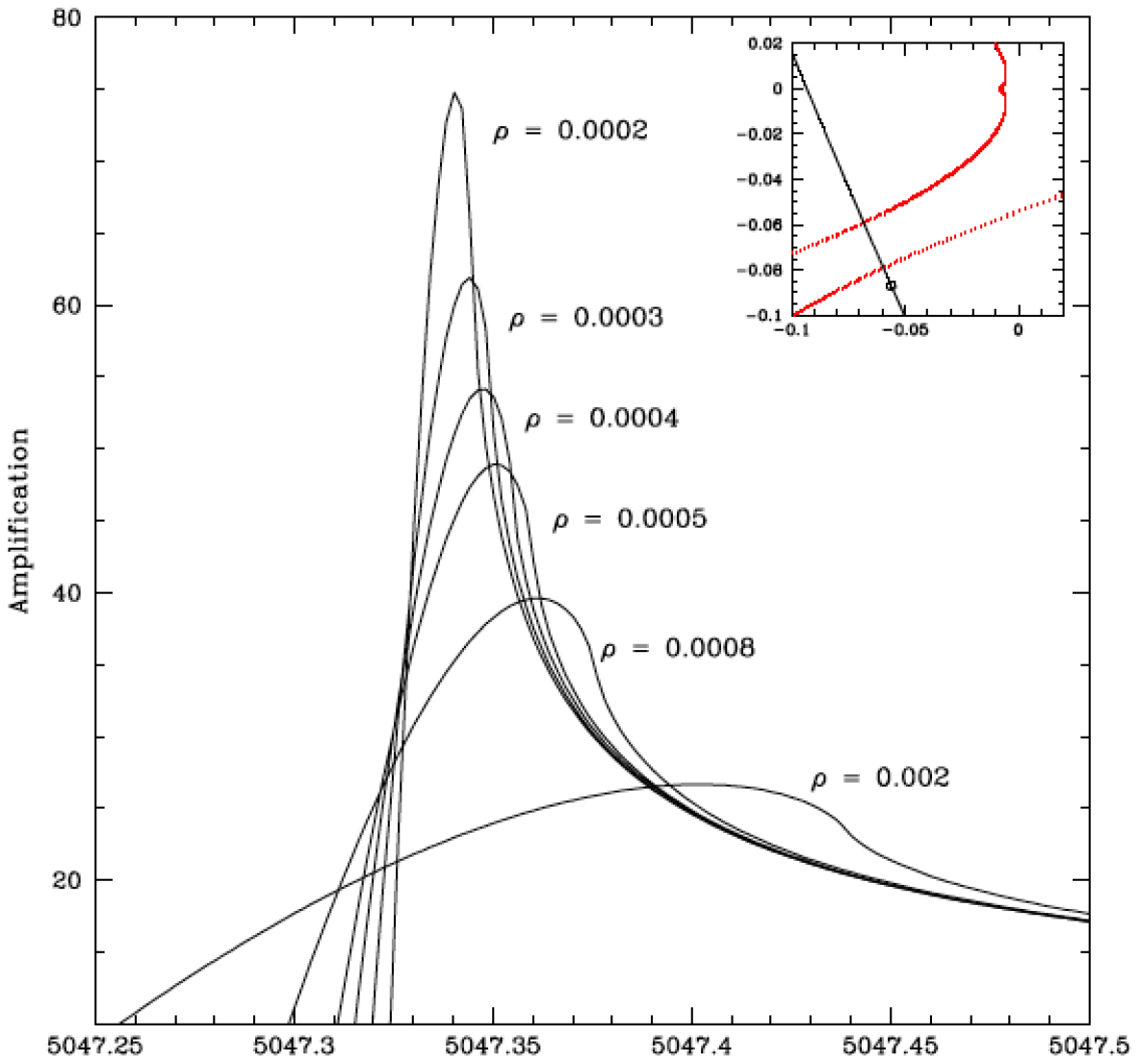}
\caption{Zoom on microlensing light curves showing a caustic crossing (see the inset) with different source sizes: $\rho=0.0001$, 0.0002, 0.0003, 0.0004, 0.0008, 0.002. The lens system here has a mass ratio $q=0.013$ and a separation $d=0.914$. }
\label{fig:4}       
\end{center}
\end{figure}

When the source crosses a caustic, the finite source effects start to be noticeable as the source projected distance from the caustic is of a few source radii \citep{pejcha09}. Fig.~\ref{fig:4} gives an illustration of this effect. 

Similarly, caustic crossings enable us to determine the source radius in Einstein units, as well as its luminosity profile, i.e. limb-darkening effects on its disk. 

Formally, the source magnification can be calculated by integration over its entire surface, each point having its own magnification. Concretely, this approach is hard to implement due to divergences near the caustics. Several algorithms and techniques have been developed to solve the extended source case \citep{dominik95, bennett96, wambsganss97, gould97, gould08, pejcha09, bennett10a}. The most efficient approach seems to be the one that consists in cutting the light curve in segments as a function of the source distance to the caustics. 

Three different methods are often applied, one when the source is relatively far from the caustics, and the other for close approaches or crossings of caustics. 
The first method is based on an analytical approximation, such as the `hexadecapole' approximation \citep{gould08, pejcha09} that evaluates the magnification of 13 points well distributed on the surface of the source. 

The second one is a more complex numerical evaluation of the source magnification that integrates the images in the lens plane. One possibility is to compute an `inverse ray shooting', i.e. to throw rays from the lens plane to the source plane, in order to calculate the images/source areas ratio. Indeed, if computing the source-to-lens path of light rays is difficult, the lens-to-source path is straightforward. One just needs to apply the equations of light deflection for a given lens model that is tested, and then count the amount of rays falling in the source disk. The ratio of images/source areas gives the total magnification. By Liouville's theorem, the surface brightness is conserved, thus the density of light rays in the image plane is uniform. This method requires a dense sampling of the image plane, a grid search on a large $(d,q)$ parameter space, and an efficient algorithm to fit the other parameters $(t_0,u_0,t_{\rm{E}},\rho,\alpha)$, where $\alpha$ is the angle between the source trajectory and the lens system axis. Furthermore, triple lens modeling is often required and has been implemented in several codes, in particular those using a faster ray shooting technique, the image centered ray shooting \citep{bennett96, bennett10a}.

The third method consists in using Green's theorem to compute a 1-D mapping of the image contours, instead of 2-D \citep{dominik95, gould97, dominik98}. Each point requires to solve a fifth order polynomial, which makes individual computations time consuming, but this disadvantage is largely compensated by the faster 1-D integration. However, the task of connecting points in the lens plane in order to obtain closed contours is sometimes confusing, in particular when the source passes near a cusp. Such a code has been released to the public by Valerio Bozza in 2017, whose algorithm has been developed over the years \citep{bozza10,Bozza18,bozza2021} and is publicly available at \cite{Bozzaurl}. The algorithm has inspired great interest in the microlensing community and has initiated the development of more  open source microlensing light curve modelling software \citep{bachelet2017,Poleski2018,Ranc2018}, that use the algorithm as their core.
\subsection{Microlens parallax, $\pi_{\rm E}$}

The parallax effects count between the most remarkable and useful to solve microlensing problems. They appear as a distorsion in the light curve, due to a non uniform motion of the observer. For fairly long microlensing events ($t_{\rm{E}}\geq 1{\rm{year}}/2\pi$), the velocity of the Earth can no longer be considered as constant and rectilinear, and this induces asymmetries in the light curve called `orbital parallax' \citep[e.g][]{gould92, alcock95, mao99, smith02}. Parallax effects can also be measured when the same event is observed from different observatories. Indeed, `terrestrial' parallax can result from simulatenous ground-based observations at different longitudes, for high-magnification events monitored at very high cadence. And `satellite' parallax can be measured when ground-based observations are combined with space-based observations.

Detection of such effects yields the Einstein radius determination, projected in the observer plane, $\tilde{r_{\rm{E}}}\equiv (2R_{\rm{Sch}}D_{\rm{rel}})^{1/2}$, where $R_{\rm{Sch}}\equiv 2GMc^{-2}$ is the lens Schwarzschild radius and $D_{\rm{rel}}^{-1}\equiv D_L^{-1}-D_S^{-1}$. If in addition the Einstein angular radius $\theta_{\rm{E}}=(2R_{\rm{Sch}}/D_{\rm{rel}})^{1/2}$ is measured from finite source effects, then the lens mass can be deduced (Gould \& Loeb 1992)
 
\begin{equation}
 M=\frac{c^2}{4G}\tilde{r_{\rm{E}}}\theta_{\rm{E}}=0.1227M_\odot\left(\frac{\tilde{r_{\rm{E}}}}{1{\rm{AU}}}\right)\left(\frac{\theta_{\rm{E}}}{1{\rm{mas}}}\right).
\end{equation}

Parallax effects are generally defined by the vector $\mathbf{\pi_{\rm{E}}}$, called microlens parallax, which magnitude gives the Einstein radius projected in the observer plane,

\begin{equation}
\pi_{\rm{E}}\equiv \frac{\pi_L-\pi_S}{\theta_{\rm{E}}}:\frac{1{\rm{kpc}}/D_{\rm{rel}}}{\theta_{\rm{E}}/1{\rm{mas}}}=\frac{1{\rm{AU}}}{\tilde{r_{\rm{E}}}},
\end{equation}

where $\pi_L/1{\rm{mas}}=1{\rm{kpc}}/D_L$ and $\pi_S/1{\rm{mas}}=1{\rm{kpc}}/D_S$.
The orientation of $\mathbf{\pi_{\rm{E}}}$ is that of the lens-source relative proper motion. Knowing $\theta_{\rm{E}}$ and $\pi_{\rm{E}}$ gives a relation between the lens and source distances, $D_L=(\pi_{\rm{E}}\theta_{\rm{E}}+1/D_S)^{-1}$.

Measuring the microlensing parallax $\pi_{\rm{E}}$ and the Einstein radius $\theta_{\rm{E}}$ represents the optimal way to derive physical parameters of the lens system. As orbital and terrestrial parallaxes are only detectable for rare high magnification or long microlensing events, the idea of measuring a satellite parallax has increased importance within the microlensing community, as was first suggested by \cite{refsdal66} and \cite{gould99}. The first space-based microlensing observations were done with {\it{Spitzer}} for a Small Magellanic Cloud event \citep[OGLE-2005-SMC-0001][]{dong07}. Such observations were also conducted for a microlensing event towards the bulge \citep[MOA-2009-BLG-266][]{muraki11} using the {\it{Deep Impact}} spacecraft. 

{\it{Spitzer}} appeared to be an excellent microlensing parallax satellite being on a solar orbit at $\sim$ 1 AU from the Earth, and the microlensing community, under a pilot program led by A. Gould \citep{gould13a, gould14a, gould15}, carried out {\it{Spitzer}} observations in 2014, 2015 and 2016. These campaigns resulted in a fair amount of highly precise Earth-Spitzer microlens parallaxes, e.g. the case of an isolated star \citep{yee15}, a binary star \citep{han16a}, and several exoplanets \citep[e.g][]{udalski15, street16}. However, a recurrent obstacle encountered with {\it{Spitzer}} microlensing light curves has been the faintness of the observed targets leading to a low S/N and the fact that only fragments are being observed, with limited baseline data. 

\begin{figure}
\begin{center}
\includegraphics[angle=-90,scale=.35,trim=0 50 50 70, clip=true]{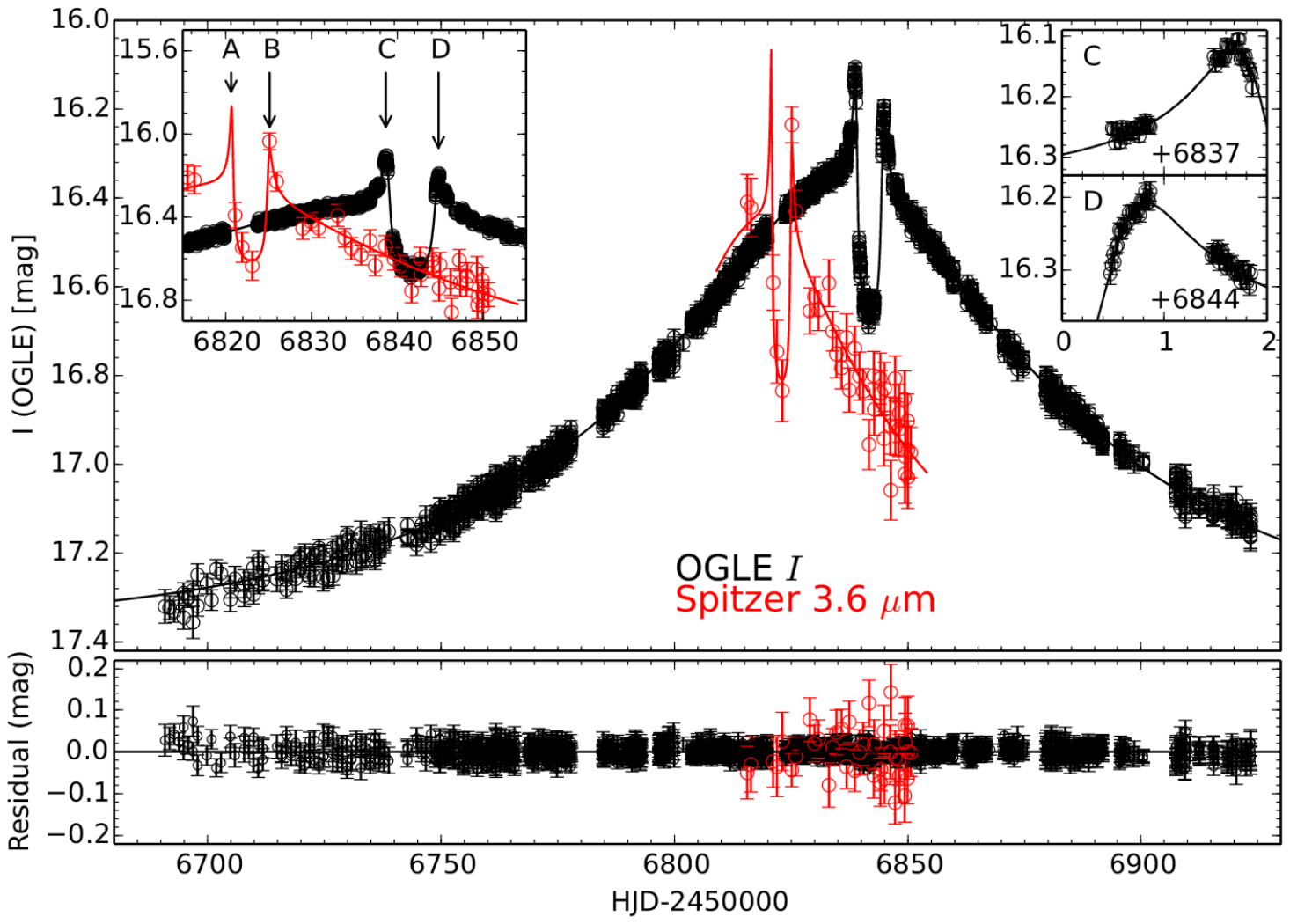}
\caption{OGLE-2014-BLG-0124Lb: a $\sim 0.5\,M_{\rm{jup}}$ planet orbiting $\sim 0.7\,M_\odot$ star, at $a_\perp\sim 3.1$ AU. The planetary anomaly and the peak of the event have been observed by {\it{Spitzer}} simultaneously to ground-based observations by OGLE. Figure from \cite{udalski15}.  }
\label{fig:5}       
\end{center}
\end{figure}

Fig.~\ref{fig:5} shows an example of a microlensing event, OGLE-2014-BLG-0124Lb, observed simultaneously with the OGLE telescope and {\it{Spitzer}}. The planetary perturbation is due to a $\sim 0.5\,M_{\rm{jup}}$ planet orbiting a $0.7\,M_\odot$ star. The two distinct light curves provided a parallax measurement, to a precision $\leq 7$ \%. Unfortunately, the error on the mass determination turned out to be much higher (30\%) due to poor constraints from finite source effects \citep{udalski15}.

Space-based microlensing observations have also been conducted in 2016 with the {\it{Kepler}} telescope in its {\it{K2-C9}} campaign \citep{howell14}, with the aim of detecting free-floating planets. These short-term campaigns have provided important insight to this mass measurement method, which will play a major role in the context of microlensing dedicated space missions like {\it{Roman}} \citep{spergel15}.

To fully exploit the potential of this effect, further investigations have been carried out to understand an underlying four-fold discrete degeneracy associated to the Earth-space microlensing parallax. This degeneracy was already discussed in the original paper by \cite{refsdal66}. Since the same microlensing event is seen from two different observatories, the two observers obtain distinct light curves with different $t_0$ and $u_0$, and the parallax is given by the difference in these quantities \citep{gould94},

\begin{equation}
{\mathbf{\pi_{\rm{E}}}}=\frac{AU}{{\mathbf{D_\perp}}}\left(\frac{\Delta t_0}{t_{\rm{E}}},\Delta u_0\right),
\end{equation}

where $\Delta t_0=t_{0,{\rm{sat}}}-t_{0,\oplus}$, $\Delta u_0=u_{0,{\rm{sat}}}-u_{0,\oplus}$, and ${\mathbf{D_\perp}}$ is the physical vector between the two observers projected in the lens plane. The direction of ${\mathbf{D_\perp}}$ gives the one of $\mathbf{\pi_{\rm{E}}}$. Since $u_0$ is a signed quantity which magnitude can be derived from the single lens light curve, but not the sign, there are actually four solutions $\Delta u_{0,\pm,\pm}=\pm|u_{0,{\rm{sat}}}|\pm|u_{0,\oplus}|$. Hence, one degeneracy creates four different orientations of the parallax vector, and a second, two different magnitudes of this vector. Several attempts have been made to break these degeneracies under special circumstances \citep[e.g][]{calchi15, yee15}, but it might actually require a revision of the parallax formalism itself. \cite{calchi16} revisited the analysis of the microlensing parallax from an heliocentric reference frame instead of the commonly adopted geocentric frame, and extended the \cite{gould94} expression of the parallax to observers in motion, as opposed to observers at rest. They showed that this new approach provides a clearer understanding of the four-fold degeneracy and interesting leverages to break it. 


In a similar way to parallax effects, if the source is animated by a significant acceleration and non rectilinear motion during the microlensing event, in other words if the source is a binary, the light curve can be affected by additional distorsions. These effects are called `xallarap' by symmetry with the parallax phenomenon. Unless the source motion coincidentally mimics the Earth's motion, these two effects are generally well distinguishable \citep{poindexter05}. Such effects have been detected in the light curve of OGLE-2007-BLG-368Lb, a cold Neptune around a K-star \citep{sumi10}.

\subsection{Orbital motion}

In the case of multiple lenses, the orbital motion of the lens components can create deviations in the light curve \citep{dominik98} due to variations in the shape and orientation of the caustics. These effects are generally negligible as microlensing events mostly involve companions with long periods. Although a Keplerian orbit would require five parameters to be characterized, microlensing light curve perturbations can only constrain two additional parameters. For a binary system, one reflects the variation of the separation $d$ between the two bodies, called $s$ here for clarity, and the second being the variation $\omega$ of the angle $\alpha$ between the source trajectory and the binary lens axis

\begin{equation}
s=s_0+{\rm{d}}s/{\rm{d}}t(t-t_0), \quad \alpha=\alpha_0+\omega(t-t_0).
\end{equation}

Considering $r_\perp=sD_L\theta_{\rm {E}}$ the projected separation of the lens system, and evaluate the instantaneous projected velocity of the lens components as the quadratic sum of $r_\perp\gamma_\perp=r_\perp\omega$ and $r_\perp\gamma_{||}=r_\perp({\rm{d}}s/{\rm{d}}t)/s$, perpendicular and parallel to the projected binary axis, respectively. Unfortunately, there is an important degeneracy in the microlensing models between $\pi_{\rm E,\perp}$ and $\gamma_\perp$ \citep{batista11, skowron11}, where $\pi_{\rm E,\perp}$ is the component of $\mathbf{\pi_{\rm E}}$ that is perpendicular to the instantaneous direction of the Earth's acceleration. 

For bright lenses, radial velocity (RV) measurements can be a good test of microlensing orbital models, as it was done by \cite{yee16} for OGLE-2009-BLG-020, confirming the model published by \cite{skowron11}.

\subsection{Mass and distance estimates}

The lens mass and distance have been directly determined in some microlensing events, thanks to microlens parallax measurements combined with Einstein angular radius and lens magnitude estimates \citep[e.g][]{bennett08a, gaudi08, batista09, muraki11, kains13, furusawa13, bennett16}. However, this is not the case for many microlensing events, and most of the time, mass and distance estimates require constraints in addition to the light curve planet-host star mass-ratio and separation.
\subsubsection{Galactic models}

A galactic model should represent the combination of the initial mass functions, stellar densities and abundances, and proper motion distribution for all types of stars \citep{duquennoy91, raghavan10, duchene13} of the galactic disk and bulge \citep{han95}. Modelling the stellar properties throughout our galaxy offers a probability distribution of the microlensing events that can occur in it. This can sometimes lead to misleading lens parameter estimates, depending on the choice of model and the simplified features it may contain. 
Most of the Galactic models used in microlensing studies were based on \citep{han95,Han2003} whose consistency with recent ground and space observations toward the bulge has not been confirmed.
\cite{Koshimoto2021} developed Galactic models, towards the galactic bulge, 
using constraints from the spatial distributions of the median velocity and velocity dispersion from Gaia DR2 \citep{DR2}, OGLE-III RC star count data \citep{nataf13}, VIRAC proper motion data \citep{Smith2018,Clarke2019}, BRAVA radial velocity data \citep{Rich2007}, and OGLE-IV star count and microlensing event data \citep{Mroz2017,Mroz2019}. Models like this can be used for all types of microlensing events and can provide valid constraints for the microlens parallax and the distance to source. 

\subsubsection{High angular resolution imaging}

Additional observations with high angular resolution imaging are strongly advocated, even for events where the microlens parallax and/or $\theta_{\rm{E}}$ were detected. Observations made simultaneously or months following the peak of the microlensing event derive constraints on the lens brightness and the baseline and thus on the lens mass with mass-luminosity relations.  It has been applied on many microlensing systems using the VLT \citep{janczak10, batista11,kubas12}, Keck \citep{sumi10, bennett14, batista14,batista15,koshimoto17} and Subaru \citep{fukui15,beaulieu16} telescopes equipped with adaptive optics (AO), or the {\it{Hubble Space Telescope}} ({\it{HST}}) \citep{dong09,bennett16}. These follow-up observations reduced drastically the uncertainties on the  physical properties of the planetary systems.

It has been shown that (AO) observations made in the decade that follows the microlensing event are able to constrain the lens mass and distance measurements with up to ten percent accuracy \citep{batista15,bennett15}. By resolving the source and lens stars, their flux ratio and relative proper motion can be deduced in addition to mass and distance constraints \citep{Beauleu2018}. 

Recently a modified version of the imaged centered ray shooting light curve modeling code of \citep{bennett96,bennett10b} has been released, where they take into account the (AO) flux of the lens and relative proper motion ensuring that the light curve fit is consistent with these constraints \citep{Bennett2023}. This allows them to resolve degeneracy of the microlens parallax or measure it in case of absence \citep{Rektsini2024,terry2024}. The code has been named \texttt{eesunhong}, in honor of the original coauthor of the code \citep{bennet2014a,bennett2014BAAS} and is now publicly available. 
\section{Detections overview}

More than 200 exoplanets have been discovered via microlensing since the first discovery 20 years ago \citep{bond04}. Although this number might seem relatively modest compared to transit and RV methods, the microlensing technique explores a population of planets that is hardly accessible, in the medium term, to other techniques. It is sensitive to a large range of planetary masses, from gas giants  down to Earth and Mars-mass planets, to wide-orbiting planets and especially planets orbiting beyond or on the snow line and is not limited by the host-star properties \citep{gaudi12}. 
Furthermore, in the right conditions, microlensing permits us to observe the influence of the gravitational field caused by the faintest objects, such as populations of free-floating planets down to Terrestial-mass \citep{Mroz2018,Koshimoto2023} and compact objects distributed around our galaxy.

The unbiased sample provided by microlensing offers a large range of planetary systems, located towards the galactic bulge or the galactic disk, with host-stars varying from main-sequence stars, to stellar remnants. This proves that this technique is ideal for constraining the demographics of both planetary and stellar objects throughout our galaxy and for providing important insights to the planetary formation theories \citep[e.g.][]{ida04,mordasini2009}. 


%

\subsection{Cold super-Earths and Neptunes}

\begin{figure}
\begin{center}
\includegraphics[angle=-90,scale=.25,trim=0 0 0 0, clip=true]{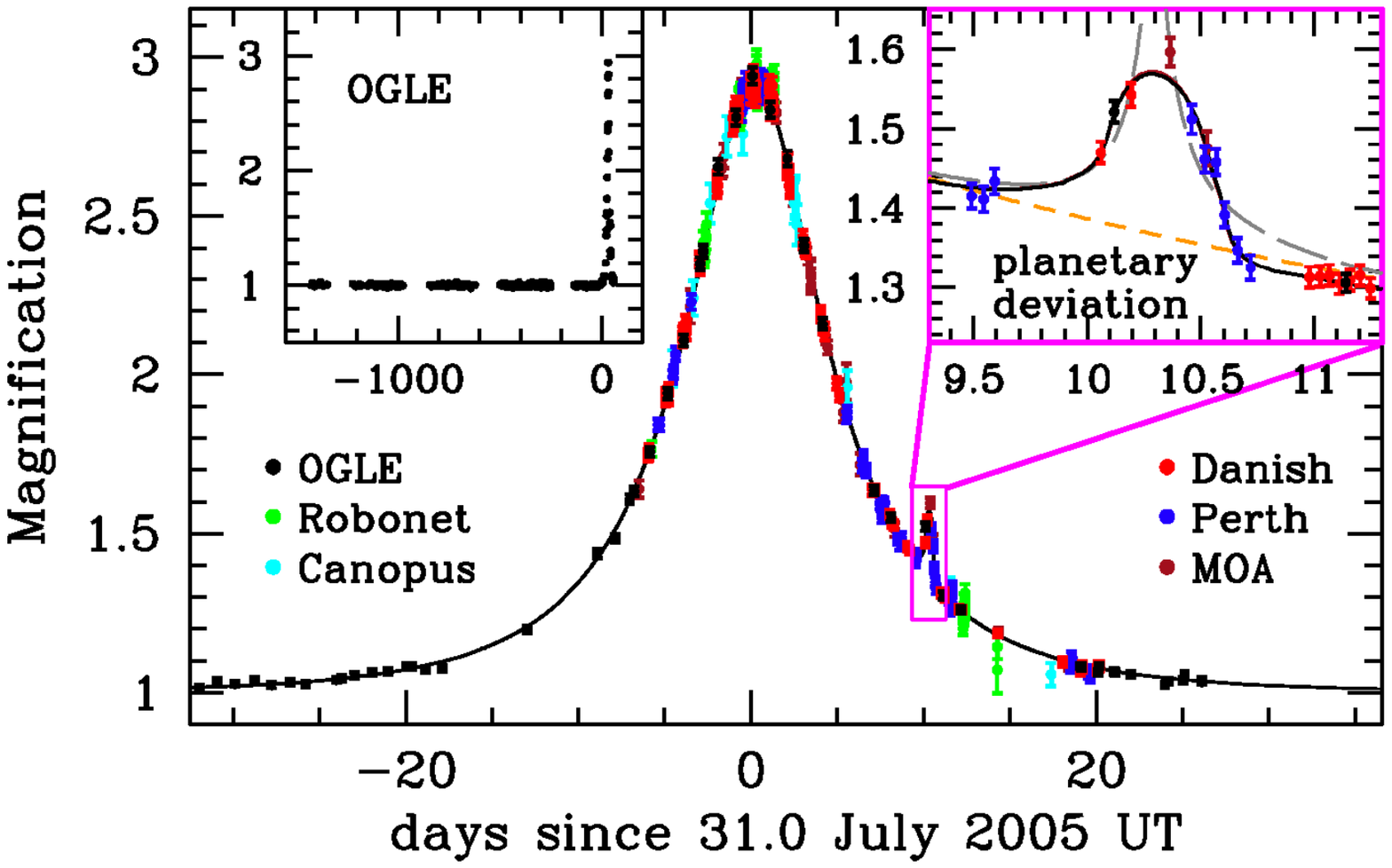}
\caption{OGLE-2005-BLG-390Lb light curve \citep{beaulieu06}. }
\label{fig:7}       
\end{center}
\end{figure}

Although the microlensing method is, like any other technique, more sensitive to massive planets, with a sensitivity decreasing with mass as $\sim \sqrt{m_p}$, the third and fourth microlensing detections were super-Earth and Neptune-like planets, suggesting that these planets are more common than gas giants. 
OGLE-2005-BLG-390Lb was the first super-Earth detected via microlensing \citep{beaulieu06}, with light curve shown in Fig.~\ref{fig:7}. A Bayesian analysis using a galactic model, combined with a $\theta_{\rm{E}}$ measurement indicated that the planet is likely to be a $\sim 5.5\,M_\oplus$ super-Earth orbiting a $M\sim 0.22\,M_\odot$ star, at a projected separation of $\sim 2.6$ AU. The equilibrium temperature of the planet is $\sim50$ K, making it the first cold super-Earth discovered at that time. 

This discovery was immediately followed by the detection of a Neptune/Uranus-like planet, OGLE-2005-BLG-169Lb \citep{gould06}, confirming that such low mass planets should be common. 
The mass and distance estimates of this system were confirmed and refined down to a 10\% precision ($m_p=13.2\pm 1.3\,M_\oplus$) with high angular resolution observations, taken 6 and 8 years after the event using {\it{HST}} \citep{bennett15}, and the Keck telescope in Hawaii \citep{batista15}, respectively. It is the first case of a planetary microlensing event for which the source and the lens are resolved, being separated by $\sim$ 62 mas on the Keck images. It allowed the detection of the lens brightness and the lens source relative proper motion. The latter provides an additional mass-distance relation of the lens system. 

More than 20 cold super-Earths, Neptunes and very low planets have been detected until today, a number that has increased significantly with the beginning of the KMTNet survey and it will rise to thousands after the launching of {\it{Roman}} \citep{Penny2019}. 

\subsection{Triple lens systems}

\begin{figure}
\begin{center}
\includegraphics[angle=0,scale=.35,trim=0 130 0 130, clip=true]{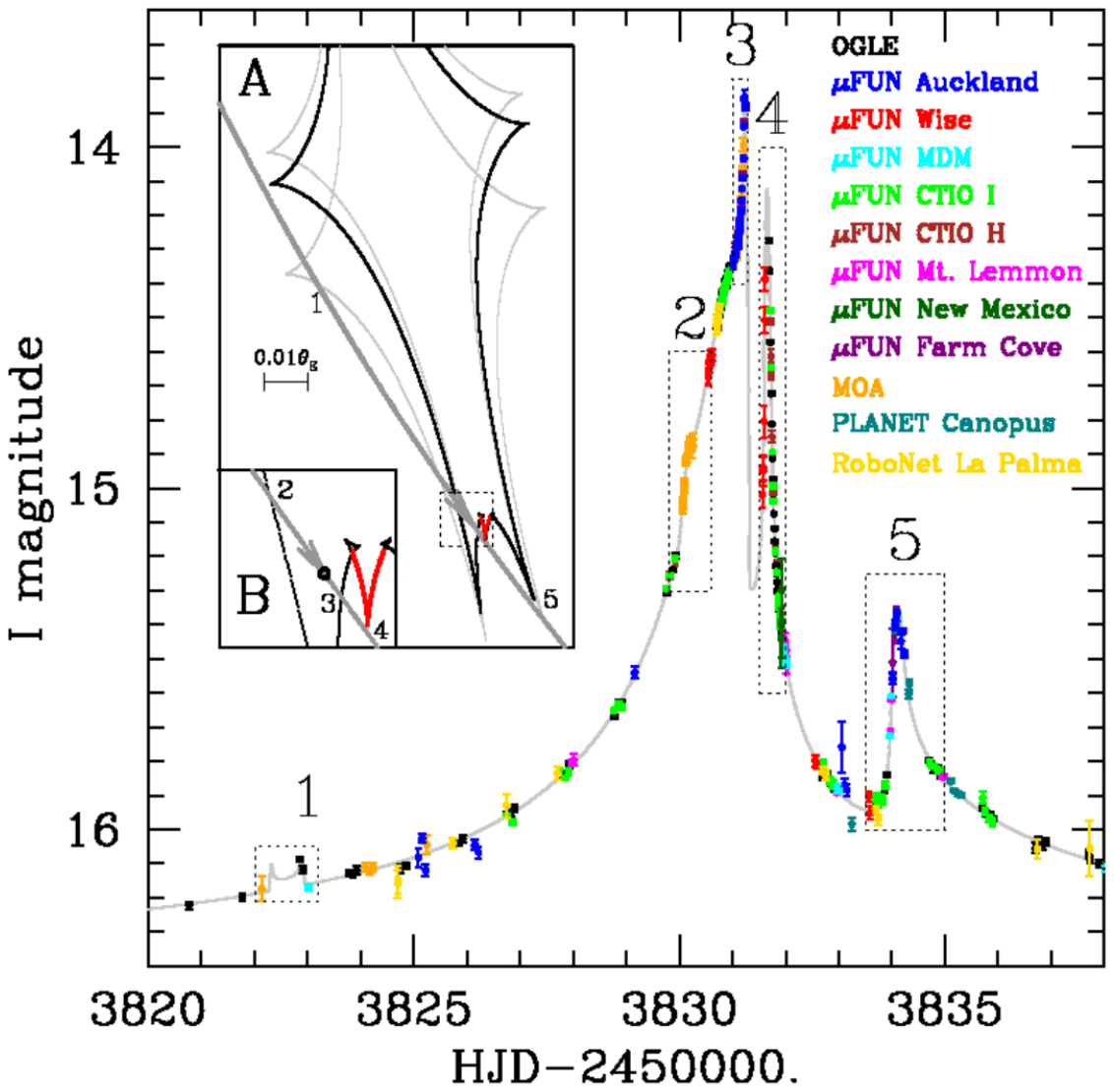}
\caption{OGLE-2006-BLG-109Lb,c: a Jupiter/Saturn analog. Figure from \cite{gaudi08}. }
\label{fig:8}       
\end{center}
\end{figure}

The first triple lens system discovered by microlensing is the 2-planet 1-star system, OGLE-2006-BLG-109Lb,c, observed in 2006 \citep{gaudi08, bennett10b}. This system presents a remarkable similarity to a scaled version of Jupiter and Saturn. It is also one of the most complex microlensing light curves observed to date, exhibiting five distinct features from the two planets (Fig.~\ref{fig:8}). Finite source and parallax effects were detected for this event, as well as orbital motion of the Saturn-like planet. This led to a complete solution to the lens system and provided the physical characteristics of the two planets. 

The next published multiple planet event was OGLE-2012-BLG-0026Lb,c \cite{han13}, composed of a Jupiter-like planet and a sub-Saturn orbiting a solar-mass star. This result was confirmed and refined by \cite{beaulieu16} who detected the flux coming from the lens using the Keck telescope.

\begin{figure}
\begin{center}
\includegraphics[angle=-90,scale=.4,trim=50 100 50 100, clip=true]{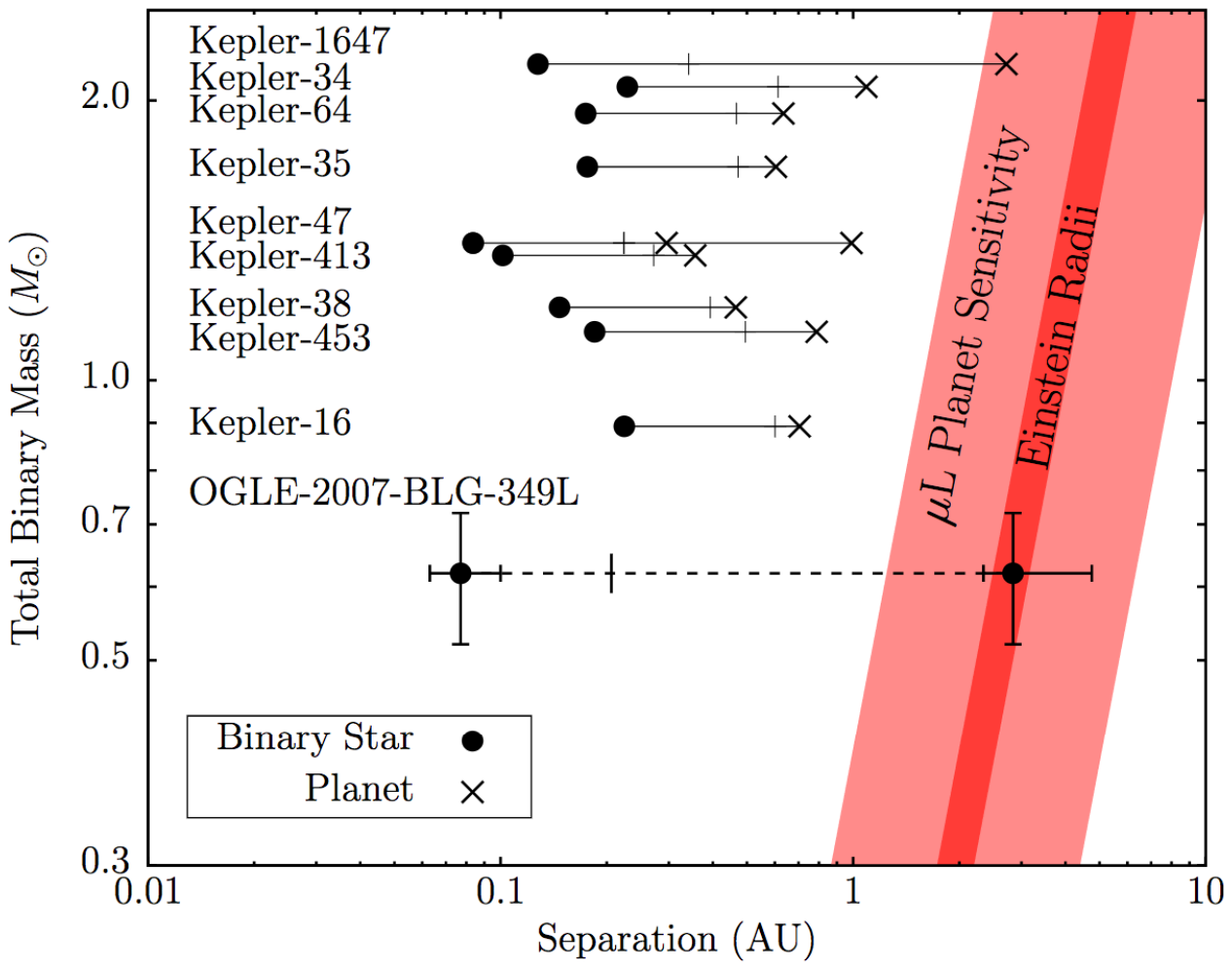}
\caption{OGLE-2007-BLG-349(AB)c: the first circumbinary planet found via microlensing compared to known circumbinary planet systems. The black dots show the orbital separation of the stars, and the crosses represent the planet separation from the stars center of mass (except for the microlensing one that shows a black dot with error bars). Figure from \cite{bennett16}. }
\label{fig:9}       
\end{center}
\end{figure}

OGLE-2007-BLG-349(AB)c is the first circumbinary planet found by microlensing \cite{bennett16}. The main features of the light curve are two cusp approaches that reveal the presence of a companion with a Saturn-to-Sun mass ratio, but only a third body could explain the slope between the two bumps. At that point of the analysis, it is not possible to disentangle the scenario of an additional planet, or an additional star. {\it{HST}} high resolution observations ruled out the 2-planet scenario because the lens flux could not explain the brightness of the host star in such a configuration. The system is then likely to be a Saturn-like planet orbiting two tight M-dwarfs, separated by only $\sim$0.08 AU, at $\sim$2.6 AU. Fig.~\ref{fig:9} compares this system to the known circumbinary systems, making it the most compact regarding the binary stars, and the largest regarding the planetary orbit.

OGLE-2013-BLG-0341Lb,c is a triple lens system with a very small planetary signal, and a large signature from the crossing of a resonant caustic due to a binary star \citep{gould14b}. A low amplitude bump in a very early phase of the event enabled the authors to identify the binary as being in wide-orbit, separated by $\sim$15 AU. The planet orbits one of them at $\sim$1 AU. It is also one of the smallest microlensing planets discovered to date ($\sim1.66\,M_\oplus$). 
A similar system has been published by \cite{poleski14}, OGLE-2008-BLG-092, i.e. a wide stellar binary in which the primary star is orbited by a Uranus-like planet ($\sim4\,M_{\rm{Uranus}}$).   

All these triple lens events (except the last one) exhibited finite source and parallax effects, allowing to derive precise estimates of their physical properties from the best models.
For the last 5 years KMTNet has been conducting a project revisiting all microlensing events that contain anomalous light curve behaviour that could not be explained by the standard 2L1S or 1L2S models. So far they have discovered six microlensing planets orbiting binary systems  along with other 14 events \citep[][and references therein]{Han2024} that could be explained by 2L2S, 3L1S and a potential five-body model \citep{Han2021}.


            
\subsection{Massive planets around very low mass stars/BDs}

Determining the frequency of planets orbiting low mass stars is of interest because these systems provide important tests to planet formation theories. In particular, the core accretion theory predicts that gas giants should be less common around low mass stars \citep{laughlin04, ida04}. Indeed, in smaller protoplanetary disks, hydrogen and helium might dissipate before the $\sim 10\,M_\oplus$ planet cores grow enough to trigger rapid accretion. This lack of efficiency in the accretion process would be responsible for a larger population of failed gas-giants around low mass stars. Numerous microlensing discoveries seem to confirm these expectations, being cold 10-40$\,M_\oplus$ planets around M-dwarfs \citep[e.g][]{gould06, muraki11, furusawa13, sumi16}. However, microlensing discoveries also revealed a population of giant planets $>1M_J$ around very low mass stars ($<0.2\,M_\odot$) \citep[e.g][]{batista11, jung15, han16b} and even around brown dwarfs (BDs) \citep{han13, choi13}. For these particular systems, it is still not clear if they have been formed through gravitational instabilities \citep{boss06} or core accretion. Moreover, the microlensing BD systems present an unusual morphology compared to most of those detected with other methods, since they are very tight binary systems ($<1$ AU versus $\sim$10-45 AU). Here again, it is not clear if they result from the same formation mechanisms as the wide systems, that are believed to come from the same process as gravitational fragmentation for stellar binaries \citep{lodato05}. 

\subsection{A jovian analogue orbiting a white dwarf star}

One of the characteristics that make the gravitational microlensing unique is its lack of host-star luminosity dependence, which permits observers to discover exoplanets orbiting any type of star, even  non-luminous objects (eg. white dwarfs). \cite{Blackman2021} reported the non-detection of a lens with most probable at least 0.5 solar masses hosting a jovian analogue planet. This is the case of MOA-2010-BLG-477Lb \citep{Bachelet2012}. 
AO follow-up observations were made for this target as part of the NASA Keck Key Strategic Mission Support (KKSMS) program \citep{Bhattacharya2018} which is
a pathfinder project for the {\it{Nancy Grace Roman Space Telescope}}. 

\begin{figure}
\begin{center}
\includegraphics[angle=0,scale=.45,trim=0 0 0 0, clip=true]{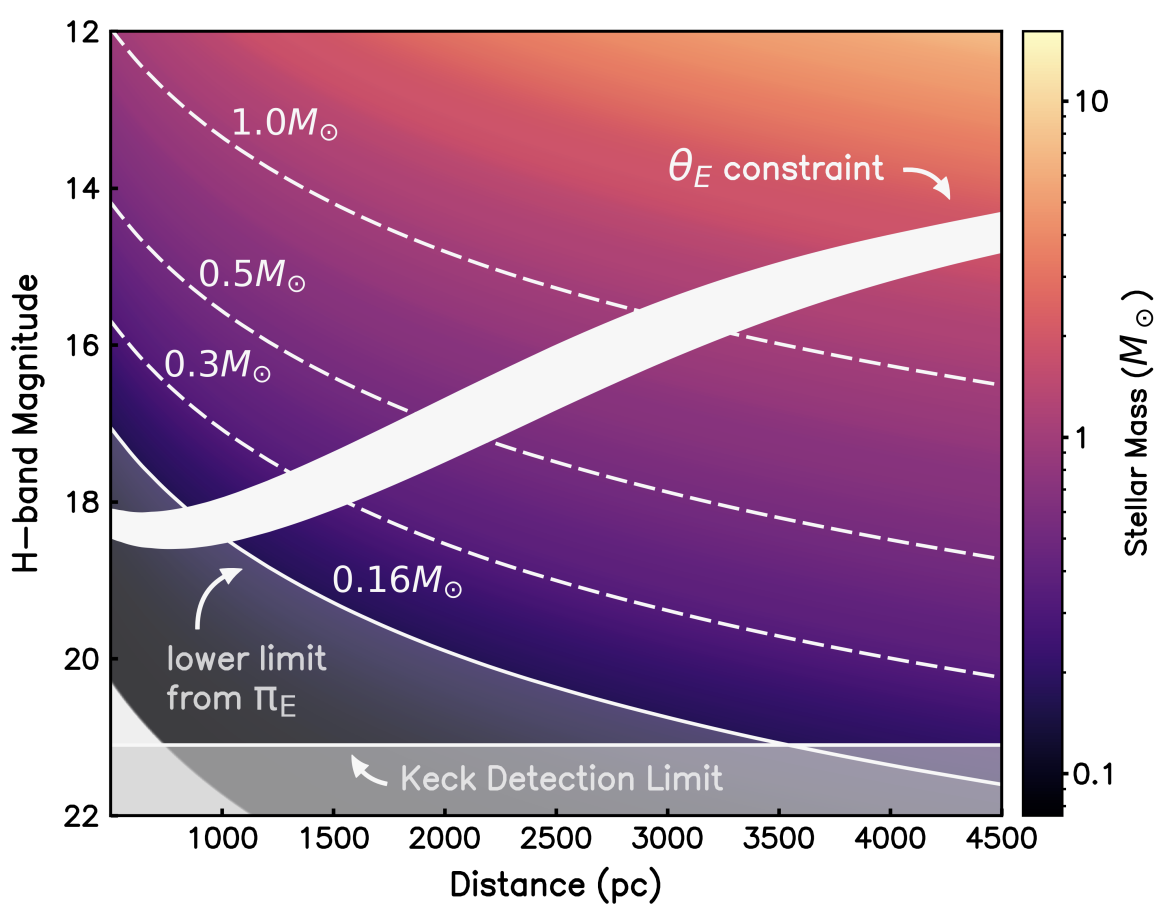}
\caption{H-band brightness of possible main-sequence host lenses. In white is the Einstein ring radius constraint,
$\theta_E$, derived from finite source effects in the event light curve. This constraint indicates that if the planet host star was main sequence, it should be visible with Keck adaptive optics as the entire area lies above the detection threshold of the H-band
images. Figure from \cite{Blackman2021}.
\label{fig9}}
\end{center}
\end{figure}

Both epochs made with the (NIRC2) instrument on the Keck-II telescope showed no sign of the lens, thus the lens magnitude should be lower than the Keck detection limit in H-band (H$\sim$21.1) as shown in Fig.\ref{fig9}. 
They estimate that the system contains a white dwarf or a rare low-mass neutron star of M = 0.53 $\pm$ 0.11 $M_{\odot}$ hosting a Jovian planet with M$_p$ = 1.43 $\pm$ 0.30 M$_{Jup}$ at a distance D$_L$ = 1.99 $\pm$ 0.35 kpc. If this non-detection proves to be valid, this would be a good example of what could be the long future of our solar system. 

\subsection{Images of rotating gravitational arcs with interferometry}

A benefit of interferometry method is that it is not constrained by the luminosity of the microlens or its passage from a caustic. In practice, when a microlensing event occurs the source image is split into several images separated by an order of microarcsecons to milliarcseconds, forming the Einstein ring (in perfect alignment). The resolution of images with such a small separation is impossible with classical telescopes, but it can be achieved by long-baseline interferometers. 
\cite{Delplancke2001} first showed that the microlensed images can cause fringe visibility patterns in interferometric observations. Several studies have been made ever since to better understand this type of observations, the likelyhood of caching an event and how to combine these observations with the microlensing models and lens parameters \citep{Dalal2003,Rattenbury2006}. \cite{Cassan2016} established a new method to connect the microlensing images positions and the fringe visibility patterns observed by long-baseline interferometers by defining a microlensing interferometric plane. Further mathematical details regarding this formalism can be found in C. Ranc's PhD thesis \cite{Ranc2015}.

\begin{figure}
\begin{center}
\includegraphics[angle=0,scale=.35,trim=0 0 0 0, clip=true]{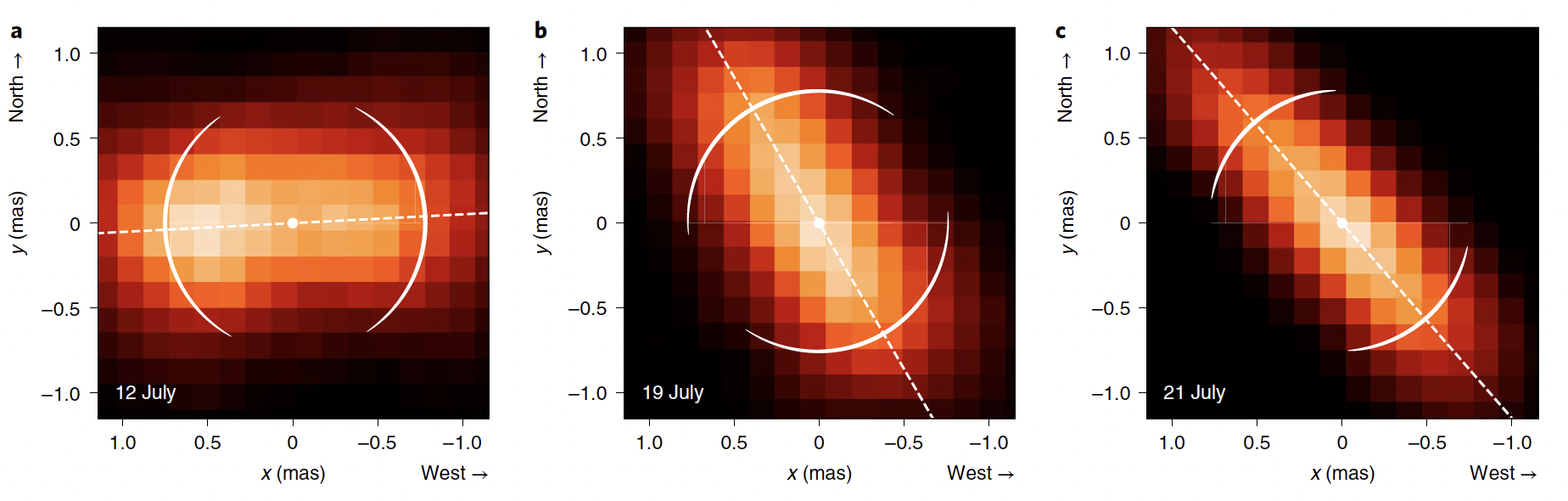}
\caption{Gravitationally lensed arcs in rotation around Gaia19bld’s microlens. From left to right, the red-scale patterns show the model-independent
interferometric reconstructions of the microlensed images for each of the three epochs. The white dot mark the position of the (unseen) microlens, and the dotted lines join the centre of the arcs. Figure from \citep{Cassan2022}.
\label{fig10}}
\end{center}
\end{figure}

This study permitted them to build observing strategies that lead to a successful real-time observation of rotating gravitationally lensed arcs of the event Gaia19bld \citep{Cassan2022,Rybicki2022}. The observations were made with the PIONIER instrument at the Very Large Telescope Interferometer. They collected five sets of interferometric data in total, one during the first epoch, before the peak and another two sets for each epoch after the peak. They used the four 1.8-m Auxiliary Telescopes at medium baseline configurations before the peak and at large baseline configurations following the peak. Using SQUEEZE \citep{Baron2010} to reconstruct the source images they discovered that Gaia19bld's arcs were captured in rotation around the lens (Fig.\ref{fig10}), the first image of this phenomenon ever observed. 

Each of the three epochs provided independent measurements of ${\it{\theta_{\rm{E}}}}$ and its direction and thus deduced a microlensing mass, M = 1.147 $\pm$ 0.029 M$_{\odot}$. This measurement is also in agreement with the mass derived by \cite{Rybicki2022} where they perform a photometric analysis of the event using Gaia and Spitzer data. 

This remarkable observation, in addition to the interferometric resolved images of the microlensing event Kojima-1Lb (TCP J05074264+2447555) \citep{Dong2019} proves that interferometric microlensing has an important role to play in exoplanet demographics and the existence of stellar-mass black holes in our Galaxy. 

\subsection{Stellar remnants}

Aside from searching for exoplanets, microlensing is also one of the rare ways to explore the population of compact objects in our galaxy. Three black hole candidates had been found in the past, MACHO-98-6, MACHO-96-5, MACHO-99-22, but they could not be confirmed because there was no measurement of $\theta_{\rm{E}}$ \citep{bennett02, poindexter05, mao02}.

\cite{shvartzvald15} analyzed a binary microlensing event, OGLE-2015-BLG-1285La,b, that was observed simulatenously from space and Earth, during the 2015 {\it{Spitzer}} campaign. They found that the lens system is likely to lie in the galactic bulge, with the more massive component being $>1.35\,M_\odot$, either a black hole or a neutron star. The authors advocate future high resolution observations of this system, in order to measure the proper motion of the bright companion.

From the OGLE-III database of 150 million objects observed from 2001 to 2009, \cite{wyrzykowski16} published a list of 15 stellar remnants candidates, combining parallax and brightness measurements from microlensing light curves. The most massive black hole of their sample is estimated to be $8.3\,M_\odot$ at a distance of 2.4 kpc. 

Recently, \citep{Mroz2024a,Mroz2024b} combined the data from the OGLE-III and OGLE-IV phases from 2001 to 2020 collecting in total nearly 20 years of OGLE observations of 62 million stars and searched for microlensing events towards the LMC. Their goal was to use a long enough dataset that will enable them to discover massive lenses, such as primordial black holes, that cause magnified events with long timescales of up to 10 years. They discovered 13 microlensing events, from which 
the longest timescale was 165.941$^{+87.842}_{-49.125}$ days. The absence of light curves with large timelength leads them to the conclusion that all the events can be described as stellar objects in the Milky Way disk or inside the LMC and find no evidence of dark matter in the form of compact objects. The authors advocate future high resolution observations could constrain the physical parameters of these lenses and reveal their precise location. In addition, their measurement of the optical depth towards the LMC is in good agreement with that from \cite{wyrzykowski2011} and confirm the low value of the optical depth from EROS \citep{Tisserand2007}.

In addition to photometric searches, recent projects have also sought to detect stellar remnants, in particular single black holes and neutron stars and even probe for the existence of intermediate-mass black holes in globular clusters \citep{kains16}, using astrometric microlensing with {\it{GAIA}} and {\it{HST}}. Indeed, in addition to the well-known photometric effect, a microlensing event also leads to a small apparent change in the position of the source \citep{hog95, dominik00}. This effect is small, but if detected, leads to direct constraint on $\theta_{\rm{E}}$, facilitating direct mass measurements via microlensing. 

\cite{Wyrzykowski2023} reported and described the properties of the 363 microlensing events included in the third Gaia Data Release. The majority of these events is located towards the Galactic bulge, with the density of events following the stellar density in the surrounding Galactic regions. They also found a low rate of individual events detected in very high Galactic longitudes as far as the Galactic anticentre. After cross-matching these events with previous studies from the OGLE database they confirmed the detection of 90 new events.


In the future the combination of ground-based and space-based surveys (e.g., {\it{GAIA}}, {\it{Euclid}} and {\it{Roman}}) will offer enhanced astrometric precision, allowing us to constrain the demographics of stellar remnants in the Milky Way.

\subsection{First confirmation of an isolated stellar-mass black hole}
Our galaxy is expected to host around $10^8$ isolated black holes but their observation can't be made by classical optical methods. Gravitational microlensing is able to undertake such an observation but its confirmation can present unfamiliar systematic errors. OGLE-2011-BLG-0462/MOA-2011-BLG-191 is a magnification event caused by a stellar black hole that proved to be a big challenge for the microlensing community. In 2022 two independent teams  released two different papers announcing the results of the photometric and astrometric treatment of the event. 

\cite{Sahu2022} used both OGLE and MOA photometric in addition to HST observations made during 2011-2O17. \cite{Lam2022} used only the OGLE data and the same astrometric data with additional HST data from 2021. The two teams realised the data treatment and modelling using different methods and concluded in two very different mass lens values, M = $7.1 \pm 1.3$ M$_{\odot}$ for \cite{Sahu2022}, indicating a stellar-mass BH and M = 2.15$^{+0.67}_{-0.54}$ M$_{\odot}$ for \cite{Lam2022}, indicating a neutron star remnant. 

Later in the same year \cite{Mroz2022} released a study where they resolve the systematic errors causing this mass difference and after deducting an extended data treatment conclude to a mass of M = $7.88 \pm 0.82 $ M$_{\odot}$.
Finally, a re-analysis of the event was conducted by \cite{Lam2023} , using the re-reducted OGLE data and an updated HST data treatment with additional observations from 2022 and confirm that the lens is a stellar BH of M = 6$^{+1.2}_{- 1.0}$ M$_{\odot}$.

\section{Exoplanet mass-ratio function}

More than 5587 planets have been discovered to date, in about 4150 planetary systems. The vast majority of them were detected by the transit, RV surveys and direct imaging i.e. detection methods that are most sensitive to close-in planets. From these surveys, \cite{winn15} estimated that half of the Sun-like stars should host a small planet (Earth to Neptune) within 1 AU, and 10\% of them a giant planet within a few AU. Although only about 200 microlensing planets have been discovered, this technique is complementary to others because it probes a different parameter space, being sensitive to cold low-mass planets beyond the snow line, at $\sim$1 to 10 AU from their host star \citep{bennett08b, gaudi12}.    

Several statistical microlensing studies have been published, before and after the survey groups had updated their observational strategy and moved to high-cadence monitoring with wide field imagers. The first studies were done even before the first planet was discovered, computing detection efficiency on microlensing events, to derive upper limits on the frequency of giant cold planets \citep{gaudi02, snodgrass04}. 
Then, from the ten first planetary detections, \cite{sumi10} analyzed the distribution of planet mass ratios beyond the snow line, and found that it could follow the power law $dN/d\log q=q^{-0.68\pm 0.20}$, implying that Neptunes would be much more common than Jupiters. 
 
\cite{gould10} used the argument of a non-biased systematic monitoring of high-magnification events (HME), to derive statistics from a sample of the 13 observed HME at that time, that involved 6 planets. They found a $\sim$36\% absolute frequency of planets, for a mass-ratio centered on $q=5\times 10^{-4}$ and a projected separation beyond the snow line ($\sim3a_{{\rm snow}}$). 

\cite{cassan12} published an analysis based on the PLANET follow-up data during 6 years, and combined their estimate with the previous results. They found a planet frequency that can be described by the power law $d^2N/d\log M\,d\log a\sim 0.80(M/M_\oplus)^{-0.65\pm 0.34}$, being consistent with every star hosting a super-Earth on average. 

\cite{clanton16} searched for a common power law to connect the results from five independent surveys, microlensing \citep{gould10, sumi10}, RV \citep{montet14} and direct imaging \citep{lafreniere07, bowler15}. They derived a planet distribution of the form $d^2N_{\rm{pl}}/(d\log m_p d \log a)=\mathcal{A}(m_p/M_{\rm{Sat}})^\alpha (a/2.5\,AU)^\beta$, where $\alpha\sim -0.57$, $\beta\sim 0.49$, $\mathcal{A}\sim 0.26\,{\rm{dex}}^{-2}$, and with an outer cutoff at $a_{\rm{aout}}\sim 9.7$.

All the statistical results presented above were based on a relatively small number of microlensing events, and $\leq$10 planets. Within the last decade, the number of microlensing events has substantially increased due to the new generation of high-cadence wide-field surveys (OGLE, MOA, KMTNet), and these planet abundances have been updated recently. \cite{shvartzvald16} estimated the planet mass ratio function from a 4-year data survey of a 8 deg$^2$ galactic field by OGLE, MOA, and Wise. Their results show that 55\% of the stars host at least one planet beyond the snow line, with a mass-ratio $-4.9<\log(q)<-1.4$, with Neptunes being $\sim$ 10 times more common than Jupiters. 

\begin{figure}
\begin{center}
\includegraphics[angle=0,scale=.3,trim=0 0 0 0, clip=true]{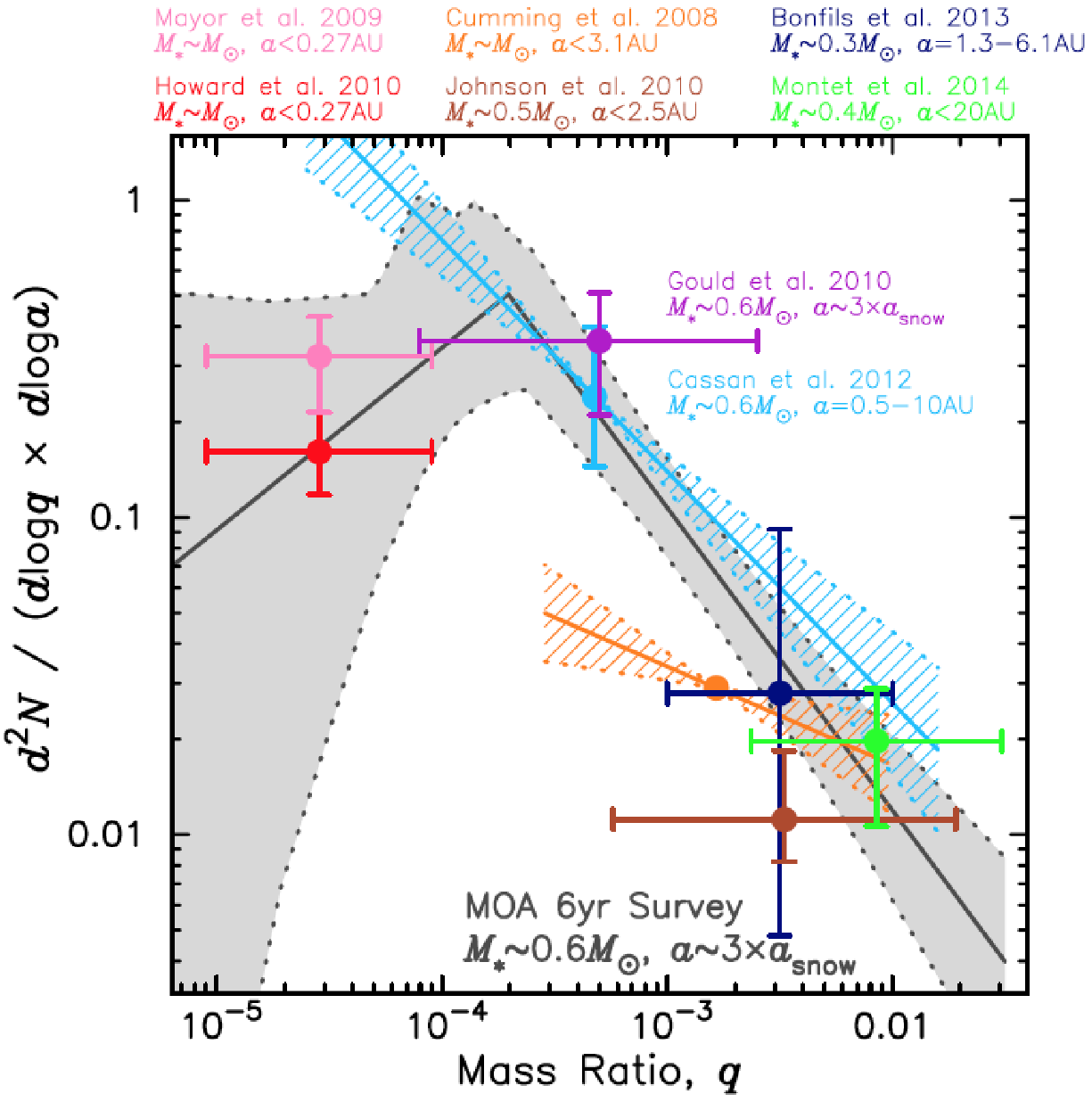}
\caption{Planetary mass ratio distribution from the 2007-2012 MOA-II microlensing survey, compared to previous microlensing and RV results
\citep{mayor09,howard10,cumming08,johnson10, bonfils13, montet14}.
Figure from \cite{suzuki16}.}
\label{fig:10}       
\end{center}
\end{figure}

\cite{suzuki16} conducted a statistical analysis over 6 years of the MOA-II microlensing survey, being the study that covered the biggest amount of microlensing events (1474) and planetary detections (23) to that time. They find that the power law describing the planetary mass-ratio function is no longer a single power law as previoulsy published, but a broken power law of the form $d^2N/(d\log q\,d\log s)=\mathcal{A}(q/1.7\times 10^{-4})^{n}s^{m}{\rm{dex}}^{-2}$, where $\mathcal{A}=0.61^{+0.21}_{-0.16}$, and $m=0.49^{+0.47}_{-0.49}$. For the low mass-ratio regime, $q<1.7\times10^{-4}$, $n=0.6^{+0.5}_{-0.4}$, and for the bigger mass ratios, $q>1.7\times10^{-4}$, $n=-0.93\pm0.13$. Their distribution thus reaches a peak around the Neptune mass ratio, as shown in Fig.~\ref{fig:10}, with a slightly lower abundance of Earth-like planets than Neptunes, beyond the snow line. They also compared their microlensing mass ratio function to previous RV results. For low mass ratios, $q\sim 3\times 10^{-5}$, it is consistent with \cite{mayor09} and \cite{howard10}, the latter being for period orbits $<$ 50 days. For bigger mass ratios, $\lesssim 10^{-3}$, their values are higher than \cite{cumming08}, and their function rises more steeply towards Neptune mass ratios. That might indicate that there are more Neptunes beyond the snow line than at $\sim$1 AU, and that cold Neptunes would not be highly subject to migration. Recent studies \citep{Silva22} seem to reinforce the findings of \cite{suzuki16} and will be included in the upcoming statistical analysis of cold exoplanets detected by the MOA-II survey.

\cite{Suzuki2018} simulated a population of planets around all the events studied  by \cite{suzuki16} using  the (IL) model \citep{ida04} and the Bern model \cite{mordasini2009}. The comparison between the planet-to-star mass-ratio distribution measured by microlensing and the theoretical predictions showed a discrepancy of exoplanet population for mass-ratios of $10^{-4}$ $\leq$ q $\leq$ 4 $\times$ $10^{-4}$. The \cite{suzuki16} observations predict $\sim$ 10 $\times$ more planets in these mass-ratio than the population synthesis models, with or without migration. However, a more complete comparison would require the planet and host star masses instead of the mass ratios, which can be derived by high angular resolution images. As part of the NASA (KKSMS) program AO follow-up observations with Keck and/or HST \citep{Bhattacharya2018} have been conducted for all the \cite{suzuki16} planetary detections. The analysis of the  \cite{suzuki16} fully sample will provide a very precise and complete planetary mass-function and will lead the way for the {\it{Roman}} mass and distance measurement techniques.

Finally, \cite{Udalski2018} conducted an analysis of seven events with mass ratios q $\lesssim$ $10^{-4}$ from multiple photometric surveys (OGLE, KMTNet, MiNDSTEp) 
and they confirm a sharp turnover in the mass-ratio function relative to that found by \cite{sumi10} and \cite{suzuki16} but at higher mass ratios. \cite{Jung2019} using 15 detected planets with KMTNet deduce a break at q$_{br}$ $\simeq$ 0.55 $\times$ $10^{-4}$, with an even more severe power-law slope but argue whether this could be caused by a drop of detection sensitivity. A new statistical analysis of more than 50 microlens planets discovered by KMTNet is being conducted, the results of which were not published yet while 
this chapter was being finalised. This will be the larger sampled analysis conducted to date, whose conclusions might challenge all the previous statistical results presented in this section. 

\section{Free-floating planets}

From the 2006-2007 MOA-II survey, \cite{sumi11} (S11 hereafter) identified an overabundance of short timescale microlensing events ($t_{\rm{E}}<2$ days) that could not be explained by known stellar and brown dwarf populations. They interpreted these light curves as signatures of either free-floating planets (FFP) or wide-orbit Jupiters (separation $\gtrsim10$ AU from their host star). \cite{bennett12} estimated that if all of this population was due to planets bound in wide orbits, their semimajor axis would likely be $>30$ AU. 

\cite{clanton17} generated a population of distant bound planets with distributions of mass and distance consistent with microlensing, RV and direct imaging surveys \citep{clanton16}, and estimated which fraction of them would not show any signature of the host star in a microlensing light curve. They tried to fit the microlensing event timescale distribution measured by S11 with a lens mass function composed of brown dwarfs, main-sequence stars, and remnants, augmented by the distribution of these bound planets with no star signature. They found that this fraction does not entirely explain the excess of short timescale events from S11, and concluded that $\sim 60$\% of it should be due to FFP. 
This would imply that FFP are $\sim 1.3$ times more abundant than main sequence stars in the galaxy.  

\cite{Mroz2017} did a more extensive analysis on the 2010-2015 OGLE-IV dataset, analysing a sample of 2617 events and found contradictory results to the S11 Jupiter-mass distribution. Their observed distribution deduces 43 less events with 0.3d $<$ $t_{E}$ $<$ 1.8d and 36 less events for 0.3d $<$ $t_{E}$ $<$ 1.3d. They did discover 6 potential candidates on shorter $t_E$ timescales ($t_{E}$ $<$ 0.5) indicating the existence of Earth-mass unbounded planets which is much more possible due to dynamical interactions in young systems \citep{Ma2016}. 

The main challenges of detecting FFP candidates is the short timelength of the event due to the small lens mass and the potential absence of finite source effects observations. These observations are required in order to measure the angular Einstein radius $\theta_E$, which is extremely crucial in this type of detections because, in contrast to $t_E$, $\theta_E$ is independent from the source-lens relative proper motion, so it immediately offers information on the mass of the lens. 

The first report of finite source effects on FFPs was made by \cite{Mroz2018}, where they used OGLE and KMTNet databases to detect a Neptune-mass free-floating planet candidate and managed to run successfully a finite-source point-lens model (FSPL) on it. More FSPL candidates were detected the years that followed \citep{Mroz2019a,Mroz2020a,Mroz2020b,Ryu2021,Kim2021,Koshimoto2023}, most of them in the regime of Terrestial-Neptune masses, which confirms the mass-function distribution estimated by \cite{Mroz2017}. 

\cite{Ryu2021} and \cite{Kim2021} were the first to study the distribution of the angular Einstein radius of FFPs using the KMTNet dataset. They discovered a gap in the distribution for 10$\mu$as $\lesssim$ $\theta_E$ $\lesssim$ 30$\mu$as, described as the "Einstein desert". This separation can also be seen in the \cite{Mroz2017} distribution. \cite{Gould2022} worked in details on explaining and demonstrating in theory the existence of this desert using the 2016–2019 KMTNet microlensing data set. 

\begin{figure}
\begin{center}
\includegraphics[angle=0,scale=.7,trim=0 0 0 0, clip=true]{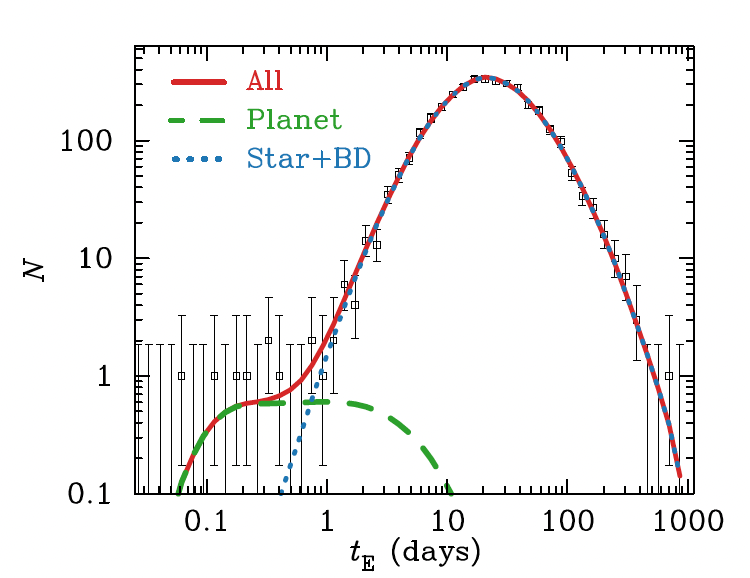}
\caption{
Timescale t$_E$ distribution based on the sample selected by \cite{Koshimoto2023}. The blue dotted line represents the known population of stars, brown dwarfs and stellar remnants, the green dashed line represents the planetary mass-population and the red line represents the best-fit model for all the populations. Figure from \cite{Sumi2023}.}
\label{fig:13}       
\end{center}
\end{figure}

\cite{Koshimoto2023} systematically analysed the MOA 2006-2014 database and selected a statistical sample of 3535 single-lens events, including six potential FFP candidates with t$_E$ $<$ 0.5 days. Out of these six candidates they found two FSPL candidates with one of them being the second Terrestial FFP candidate so far. \cite{Sumi2023} used the statistical sample selected by \cite{Koshimoto2023} and measured the FFP mass-function distribution on both $t_E$ and $\theta_E$. They confirm the non-existence of Jupiter analogues claimed by S11 and find a massive amount of lower mass FFP candidates with power law dN/dlogM = 2.18 $^{+0.52}_{-1.40}$ M$_{\odot}$. Furthermore, they compare the FFP mass-function distribution with the mass-function of bound planets, as shown in Fig. \ref{fig:13} and found that, at Earth-masses, FFPs are an order of magnitude higher than the bound planets but much fewer in Jupiter-masses. This implies that low-mass planets are more likely to be ejected from a planetary system, which confirms the findings of \cite{Mroz2017} and the predictions from the core accretion theory \citep{Ma2016}.


The three ground-based microlensing surveys have discovered in total 10 microlensing events identified as FFP candidates, with 6 of them described by Neptune-Earth mass planets. Meanwhile,  \cite{Sumi2023} suggest that {\it{Roman}} will be able to observe 988$^{+1848}_{-424}$ FFPs down to Mars-mass. Confirming whether these candidates are free-floating-planets or wide-orbit planets is the next big challenge. AO follow-up observations have been used to explore the possibility of a distant host-star existence for some of these candidates \cite{Mroz2024a} but the limiting magnitude of the observations doesn't permit a certain conclusion. Further work and deeper imaging observations will be needed to confirm all the present and future FFP candidate detections. Space missions like {\it{JWST}} and {\it{Euclid}} will most probably play a major role in this endeavor. 
\cite{Bachelet2022} has suggested  that a joint mission between {\it{Euclid}}d and {\it{Roman}} can increase the number of FFP detections and provide immediate accurate mass measurements down to Mars-masses via the microlensing parallax. 

\section{Conclusion and future prospects}

Gravitational microlensing has proven to be an excellent technique to explore the demographics of our galaxy. Its independence of the lens brightness allows it to observe any massive object residing at the darkest corners of the Milky Way, as long as sufficiently bright background sources are sampled. In the last two decades microlensing has presented a moderate but extremely diverse sample of stellar objects. It began with scientific efforts to observe dark matter in the galactic halo but along the way so much more has beed discovered. From Terrestial and Neptunian-mass planets to gas giants around extremely low mass M-dwarf stars and from N-body planetary systems, down to free-floating planets, to the possible observation of our solar-system's destiny and the confirmation of stellar-mass black holes and finally back to the search for evidence (or non-evidence) of dark matter in form of compact objects. The discovery of low mass planets beyond the snow line, with unbiased host-star sensitivity in addition to the numerous unbound planet discoveries, create complementary constraints to planet formation theories. Meanwhile, stellar remnant observations can help us understand better the stellar composition of our galaxy.  

Every new discovery forced the microlensing community to exceed previous limitations and challenge pre-existing methods and modelling techniques. The need to increase the number of exoplanet detections led to a new generation of high-cadence wide-field surveys, improved data treatment pipelines and more telescope collaborations, connecting the whole globe. It also required the possibility of real-time light-curve modelling to permit predictions of a possible anomaly appearance revealing the existence of another lens. The demand for fast and rigorous light-curve fitting triggered the interest of the community resulting in multiple modelling methods and extended light-curve studies. The use of second order effects, in particular the microlens parallax and the finite source effects lead to a breakthrough for confirming exoplanet detections and measuring their precise physical parameters. Nevertheless, the limited observed signal quality didn't always permit to deduce accurate results and further observations were needed. High angular resolution follow-up observations deducted in the decade that followed the source-lens alignment proved a strong constraint for mass and distance measurements. Furthermore, the use of AI techniques allowed the community to explore a vast number of complex and diverse light-curves and to understand better lensing degeneracies and ways to solve them.

\begin{figure}
\begin{center}
\includegraphics[angle=0,scale=.5,trim=0 0 0 0, clip=true]{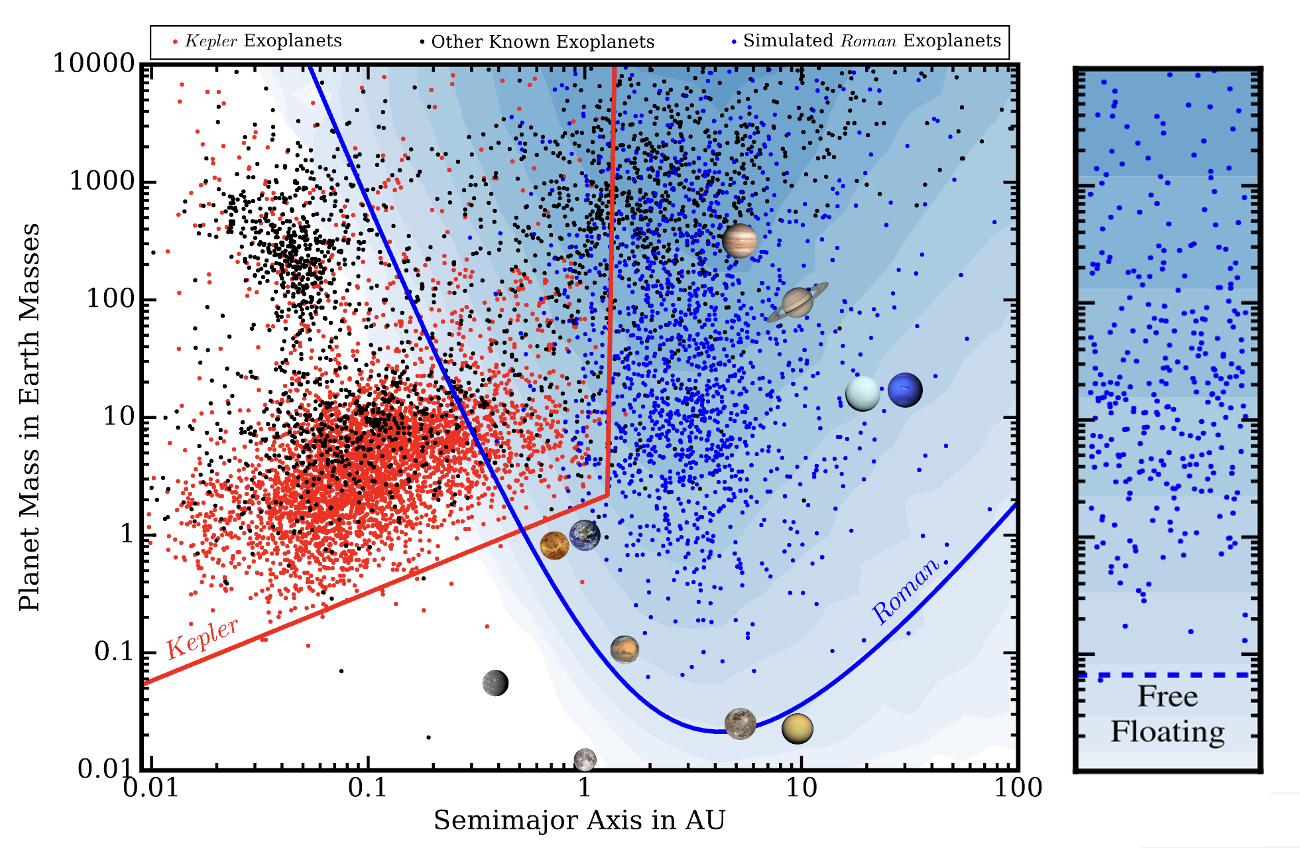}
\caption{Sensitivity range of the {\it{Roman}} mission in a planet-mass versus semi-major axis diagram \citep{Penny2019}, complementary to the parameter space probed by the {\it{Kepler}} mission. Figure credits: This research has made use of the NASA Exoplanet Archive, which is operated by the California Institute of Technology, under contract with the National Aeronautics and Space Administration under the Exoplanet Exploration Program \citep{Akeson2013}}.
\label{fig:14}       
\end{center}
\end{figure}

Starting from what seemed like an irrational idea of an improbable detection method, to the first few detections almost a decade later followed by more than twenty years of dedicated effort and hard work, microlensing is now heading to a new observational era. {\it{Roman}} is the first space mission with a dedicated survey to this technique, promising a really bright future for the microlensing community. One of the main goals of this mission is comprehending demographics of planets down to Mars mass at separations ranging from the habitable zone to hundred AU throughout the galaxy with maximum sensitivity to planets orbiting at roughly the Einstein ring radius. For solar-type stars in the Galactic disk and source-stars in the Galactic bulge, this radius is positioned at $\sim$1-10 AU, i.e. close to or beyond the snow line. Moreover, it will be possible to detect masses lower than Mercury Fig.\ref{fig:14}, and even down to the mass of Ganymede at this separation range \citep{Penny2019}. This means that in 5 years {\it{Roman}} will manage to exceed the low-mass sensitivity limit of the ground-based surveys while discovering thousand of exoplanets in a vast variety of orbital separations, something that would require decades of observations for space-based transit missions. Furthermore, it is predicted to detect hundreds of free-floating planets \citep{Penny2019,Johnson2020,Sumi2023} adding more constraints to formation theories.

As discussed in previous sections, accurate measurement of the physical parameters of the planetary system are required for comparing the statistical results of microlensing surveys with the population synthesis models and formation theories. {\it{Roman}} will be able to conduct high-angular resolution follow-up observations regularly on the events previously discovered but it takes time for the source and lens to be resolved. The {\it{Euclid}} precursor observations of the {\it{Roman}} fields can guarantee a maximum number of source-lens resolved images and thus, accurate mass measurements for all these  events \citep[][Rektsini et al in prep.]{Bachelet2022}. Furthermore, \cite{Bachelet2022} has shown that a joint mission between {\it{Euclid}} and {\it{Roman}} will ensure an even higher number of immediate accurate mass measurements for bound and unbound planets via the microlens parallax. 
In addition to space-based surveys, ground-based telescopes \citep[e.g. Rubin][]{Street2023a} will also join the "hunt" with {\it{Roman}}, either as follow-up or conducting previous observations of the Roman fields or with simultaneous observations for satellite microlens parallax detections. 

Moreover, data collected during the RGBTDS will also be valuable for detecting transiting exoplanets. It is estimated that the number of planets discovered via transit method in the {\it{Roman}} fields could reach the 200.000 \citep{Wilson2023}, the majority of them being giant planets in close-orbits but it is expected to discover a few thousands of smaller planets as well. Finally, this survey will also provide essential and groundbreaking knowledge regarding the stellar black hole demographics in our galaxy \citep{Sajadian2023} with further investigation on the existence of primordial black holes 
\citep{Fardeen2024} using astrometric microlensing and even explore the black hole populations via astrometric gravitational wave detections \citep{Wang2022}. 

Reflecting on all the discoveries of the last three decades and all the fascinating and somehow curious discoveries to come, this is without a doubt a very exciting moment for the microlensing and the astrophysical community. 



\section{Cross-References}
\begin{itemize}
\item Exoplanet Occurrence Rates from Microlensing Surveys 

\item Microlensing Surveys for Exoplanet Research (OGLE Survey Perspective)

\item Microlensing Surveys for Exoplanet Research (MOA)

\item Korea Microlensing Telescope Network
\end{itemize}

\section{Acronyms}
MACHO: MAssive Compact Halo Objects\\
EROS: Exp\'e rience de Recherche d'Objets Sombres\\
OGLE: Optical Gravitational Lens Experiment\\
MOA: Microlensing Observations in Astrophysics\\
PLANET: Probing Lensing Anomalies NETwork\\
$\mu$FUN: $\mu$lensing Follow-up Network\\
MiNDSTEp: Mincrolensing Network for the Detection of Small Terrestrial Exoplanets\\
KMTNet: Korean Microlensing Telescope Network\\
UTGO: University of Tasmania Greenhill Observatory\\
LCO: Las Cumbres Observatory\\
RGBTDS: Roman Galactic Bulge Time Domain Survey \\
ZTF: Zwicky Transient Facility \\
GAIA: Global Astrometric Interferometer for Astrophysics \\
AO: Adaptive Optics

\begin{acknowledgement}
NER would like to thank Naoki Koshimoto, Etienne Bachelet, Jean-Philippe Beaulieu and P. Scott Gaudi for their comments as well as Andrew Cole and Thomas Plunkett for proof-reading this manuscript.
V.B. would like to thank Clément Ranc for valuable discussions and Noé Kains for an input on black holes.\\
Work by NER was supported by the University of Tasmania through the UTAS Foundation and the endowed Warren Chair in Astronomy and the ANR COLD-WORLDS (ANR-18-CE31-0002). This research was also supported by the Australian Government through the Australian Research Council Discovery Program (project number 200101909) grant awarded to Andrew Cole and Jean-Philippe Beaulieu. 
\end{acknowledgement}

\bibliographystyle{spbasicHBexo}  
\bibliography{HBexoTemplatebib} 


\end{document}